%% file: RC_revised.tex
\documentclass[nofootinbib]{article}

\usepackage{graphicx}

\usepackage{rotating}

\usepackage{dcolumn}

\usepackage{amssymb}

\usepackage{amsmath}

\usepackage{bm}

\usepackage{listings}

\usepackage{fancyhdr}

\usepackage{wrapfig}

\usepackage{subfig}

\usepackage{jheppub}

\include{mydefs}

\newcommand{\C}{{\mathbb C}}
\newcommand{\R}{{\mathbb R}}

\title{Deformations of GR, Geometrodynamics and Reality Conditions}

\author{Kirill Krasnov$^1$,} 
\emailAdd{kirill.krasnov@nottingham.ac.uk}
\author{Ermis Mitsou$^2$}
\emailAdd{ermitsou@physik.uzh.ch}

\affiliation{$^1$School of Mathematical Sciences, University of Nottingham, NG7 2RD, UK}
\affiliation{$^2$Center for Theoretical Astrophysics and Cosmology, Institute for Computational Science, University of Zurich, CH--8057 Z\"urich, Switzerland}

\abstract{In four dimensions complexified General Relativity (GR) can be non-trivially deformed: There exists an (infinite-parameter) set of modifications all having the same count of degrees of freedom. It is trivial to impose reality conditions that give versions of the deformed theories corresponding to Riemannian and split metric signatures. We revisit the Lorentzian signature case. To make the problem tractable, we restrict our attention to a four-parameter set of deformations that are natural extensions of Ashtekar's Hamiltonian formalism for GR. The Hamiltonian of the later is a linear combination of $EEE$ and $EEB$. We consider theories for which the Hamiltonian constraint is a general linear combination of $EEE, EEB, EBB$ and $BBB$. Our main result is the computation of the evolution equations for the modified theories as geometrodynamics evolution equations for the 3-metric. We show that only for GR (and the related theory of Self-Dual Gravity) these equations close in the sense that they can be written in terms of only the metric and its first time derivative. Modified theories are therefore seen to be essentially non-metric in the sense that their dynamics cannot be reduced to geometrodynamics. We then show this to be related to the problem with Lorentzian reality conditions: the conditions of reality of the 3-metric and its time derivative are not acceptable because they are not preserved by the dynamics. Put differently, their conservation implies extra reality conditions on higher-order time derivatives, which then leaves no room for degrees of freedom.
}

\begin{document}

\maketitle

\section{Introduction}

Four-dimensional {\it complexified} General Relativity (GR) can be non-trivially deformed without changing its dynamical content -- the deformed theories continue to describe two propagating degrees of freedom. These theories were observed to exist following the 1986 discovery of a new Hamiltonian formulation of General Relativity \cite{Ashtekar:1986yd}, \cite{Ashtekar:1987gu}. This new formalism led to a ``pure connection" formulation of GR described in \cite{Capovilla:1989ac}, which also exhibited a one-parameter family of deformations of GR. The latter was further studied in \cite{Capovilla:1992ep}. The generalisation to an infinite number of deformation parameters was described in \cite{Bengtsson:1990qg}. Bengtsson has studied these theories extensively referring to them as neighbours of General Relativity, see e.g. \cite{Bengtsson:1991bq}. 

This infinite-parameter family of four-dimensional gravity theories was rediscovered in a Lagrangian formulation in \cite{Krasnov:2006du}, see also \cite{Bengtsson:2007zzd} for the relation to earlier work.  The more economical ``pure connection" description of the deformed theories was given in \cite{Krasnov:2011up}, \cite{Krasnov:2011pp}. This, in particular, led to a realisation that the pure connection formulation of GR (and deformed theories) works simplest for a non-zero value of the cosmological constant. In all known descriptions, the deformed gravity theories are dynamical theories of variables different from the metric. An explicitly metric formulation is possible, but complicated, and was worked out in \cite{Freidel:2008ku}, \cite{Krasnov:2009ik}, following the procedure of integrating out all the auxiliary fields  present in the action. 

The reality conditions that give rise to modified Riemannian and split signature four-dimensional gravity are trivial to impose. This is done by restricting one's attention to either ${\rm SO}(3,\R)$ or ${\rm SO}(1,2)$ connections. In contrast, to describe Lorentzian GR in this formalism one needs to work with complex ${\rm SL}(2,\C)\sim{\rm SO}(3,\C)$ connections.\footnote{This is related to the built-in chirality of the formalism, which works with only one chiral half of the relevant Lorentz group.} Reality conditions then need to be imposed to recover real Lorentzian signature metrics. This issue was surrounded by a cloud of mystery from the very first days of the deformed theories, in that it was never clear whether there is any form of the reality conditions that could render them into physical Lorentzian signature gravity theories. 

The main aim of the present paper is to revisit the issue of the Lorentzian reality conditions for deformations of GR. Our main result is that the reality condition that demands that the 3-hypersurface metric is real (together with its time derivative) is not in general an admissible condition because it is not preserved by the time evolution. One can insist on maintaining that condition by imposing the reality of the higher-order time derivatives, but this then removes all the degrees of freedom of the theory. Only for General Relativity (and a closely related theory of Self-Dual Gravity) this reality condition is admissible. This result makes it unlikely that the modified theories \cite{Bengtsson:1990qg}, \cite{Krasnov:2006du} exist as physical Lorentzian signature gravity theories. We will return to the interpretational issues in the discussion section.

Our desire is to make the description in this paper as general as possible, but not too general to loose the ability to perform explicit computations. For this reason, we restrict our attention to a four-parameter family of modified theories. This four-parameter family can be motivated as a very natural extension of Ashtekar's Hamiltonian formalism for GR. The Hamiltonian constraint of the latter, using notations that are familiar to practitioners but will also be explained below, is a linear combination of the $EEE$ term encoding the cosmological constant and $EEB$ term, the main Ashtekar Hamiltonian. We add to this mix the terms $EBB$ and $BBB$, which is certainly natural to consider. We will see that one of these terms can always be removed by a field redefinition of the sort $E\to E+B$, so there is effectively a one-parameter family of modifications that we are considering. These modifications can be motivated as the most general set of theories for which both the Hamiltonian constraint and the spatial 3-metric remain polynomial in the basic variables. All this will become more clear in the main text. 

We perform all computations having the application to Lorentzian signature and reality conditions in mind. However, many of our results are equally applicable to the cases of Riemannian and split signatures. Even for those signatures it is a non-trivial question to what extend the modified theories can be formulated as dynamical theories of an evolving 3-metric. Our computations provide a complete answer to this question, for the considered four-parameter family of modifications. Thus, the logic that we adopt will be to proceed with calculations in the complexified case as far as possible, so that the results described have the widest applicability. Only after working out the geometrodynamics interpretation of the modified theories will we return to the issue of the Lorentzian reality conditions.

\bigskip

The paper is organised as follows. We start in Section \ref{sec:action} by discussing different possible Lagrangian formulations of the modified theories. We proceed to describe the associated canonical formalism in Section \ref{sec:canonical}. We perform here the computation of the evolution equation for the 3-metric for a general modified theory, which then shows the need to specialise to some explicit examples. We introduce and analyse the minimal polynomial modifications in Section \ref{sec:polynomial}. The geometrodynamics equations are obtained in this section. We then discuss the issues related to the reality conditions in Section \ref{sec:reality}. We conclude with a discussion. There are also two Appendices, one reminding the passage from the considered here chiral formalism to the non-chiral Einstein-Hilbert metric formulation, the other relating the reality constraints on the $B$-field with the reality of the Urbantke metric.

\section{The action}
\label{sec:action}

\subsection{$BF$ formulation}  \label{sec:BF}

The BF formulation of gravity theories is described in many sources, see e.g. \cite{Celada:2016jdt}, so we will be brief. Our starting action takes the form 
\beq \label{eq:Sh}
S_h := \int \[ \frac{1}{\iu} \( B_i \we F^i - \frac{1}{2} \, \psi^{ij} B_i \we B_j \) - \ph \cH(\bm{\psi}) \] \, .
\eeq
Here 
\beq \label{eq:Fdef}
F^i := \ed A^i + \frac{1}{2}\, \vep^{i}{}_{jk} A^j \we A^k \, ,
\eeq
are the curvature 2-forms of a complex ${\rm SO}(3,\Cs)$ connection 1-form $A^i$, the $B_i$ are complex 2-forms, $\psi^{ij}$ is a symmetric matrix of complex 0-forms, $\ph$ is a complex 4-form, $\cH$ is an ${\rm SO}(3,\Cs)$ invariant matrix function of $\psi^{ij}$. The indices $i,j,k,\ldots = 1,2,3$. The objects $\vep^i{}_{jk}$ in the curvature are the ${\mathfrak so}(3)$ structure constants.

We have placed the factor of $(1/\iu)$ in front of the action. As we already discussed in the Introduction, our desire is to present our results in as general form as possible, i.e. applicable to all possible metric signatures. In the Riemannian and split signature settings all fields need to be taken to be real from the start, and no factor of the imaginary unit in the action is necessary. On the other hand, in the setting when all fields are complex-valued, this factor can be absorbed into the fields. However, it is with the factor as introduced that the Lorentzian signature reality conditions take their most natural form. We will thus keep this factor in the action, as our final aim is to clarify the Lorentzian case reality conditions. It should be kept in mind that all the factors of the imaginary unit that appear in our formulas stem from this factor being introduced in (\ref{eq:Sh}), and that this factor is absent in the Riemannian and split signature cases. This could have been easily dealt with by introducing an object $\iu$ whose square is $\iu^2=\sigma$, with $\sigma=\pm 1$ depending on the signature desired. We, however, proceed with the usual $\iu^2=-1$ to not overburden the notation. An interested reader will easily find a way to modify all the formulas that follow to make them applicable to the Riemannian and split signatures. 

Let us introduce the following notations for the matrix of partial derivatives of $\cH$ and its inverse
\beq \label{eq:defHij}
\cH_{ij} := \frac{\pa \cH}{\pa \psi^{ij}} \, , \qquad \cH^{ij} : =(\bm{\cH}^{-1})^{ij}.
\eeq
Thus, $\cH^{ij}$ denotes the inverse matrix of $\cH_{ij}$. Similarly, we will use $\psi_{ij}$ to denote the inverse matrix of $\psi^{ij}$. Later on we will see that $\cH_{ij}$ behaves as an ``internal'' metric. Despite this, we will not use it to displace indices from their natural positions to avoid confusion during contractions. We will always explicitly indicate the metric used to perform a contraction, even when this is the Killing form of the gauge group $\de_{ij}$ (or $\et_{ij} := {\rm diag}(-1,1,1)$ in the split signature case). The equation of motion of the Lagrange multiplier $\psi^{ij}$
\beq \label{eq:EOMpsi}
\frac{\iu}{2}\, B_i \we B_j = \ph \cH_{ij} \, ,
\eeq
makes $B_i$ related to the ``square root'' of the internal metric. 

The case of General Relativity  corresponds to an affine $\cH$ function 
\beq \label{eq:HGR}
\cH_{\rm GR}(\bm{\psi}) = \la + \psi^{ij} \delta_{ij} \, ,  \hspace{1cm} \la \in \Rs \, ,  
\eeq
as shown explicitly in appendix \ref{app:toEH} for the interested reader.\footnote{A more conventional description of that action is found by simply redefining $\psi^{ij} \to \psi^{ij} - \la \de^{ij} /3$, in which case the constraint imposed by $\ph$ is the tracelessness of $\psi^{ij}$. \label{ft:stdform1}} We see that in this case the internal metric is equal to the Killing form $\cH^{\rm GR}_{ij} = \de_{ij}$.
For general functions $\cH(\bm{\psi})$, the holomorphic action \eqref{eq:Sh} corresponds to the deformations of complex GR discussed in the introduction.\footnote{It should be noted that this is not the description available in the early work on the subject \cite{Krasnov:2006du}. The description in this reference is to start with the GR case as described in footnote \ref{ft:stdform1} and then generalize the cosmological constant $\la$ to a function $\la(\bm{\psi})$. The two formulations are related to each other through a redefinition of $\bm{\psi}$. The advantage of our formulation is that the $\cH$ function will become the Hamiltonian constraint in the canonical formalism.} One can integrate out $\psi^{ij}$ and subsequently $\ph$, to obtain an unconstrained $BF$ action, i.e. of the ``$BF$ + potential'' form \cite{Krasnov:2009iy}
\beq \label{eq:BFpV}
S_h \to \frac{1}{\iu} \int \[ B_i \we F^i - V(\bm{Y}) \] \, , \hspace{1cm} Y_{ij} := B_i \we B_j \, , 
\eeq
where $V$ can only be a homogeneous matrix function of degree one $V(\al\bm{Y}) \equiv \al V(\bm{Y})$ due to diffeomorphism invariance.\footnote{In fact, even GR can be put into this form by a field redefinition trick \cite{Krasnov:2017dww}.}

\subsection{Pure connection formulation} \label{sec:purecon}

We can also integrate out the auxiliary fields $B_i$ to obtain a simpler description. The equation of motion for the $B_i$ field is $F^i = \psi^{ij} B_j$, which is solved by 
\beq \label{eq:Bisol}
B_i = \psi_{ij} F^j \, ,
\eeq
where we remind that $\psi_{ij}$ denotes the inverse matrix of $\psi^{ij}$. This can then be substituted into the equation of motion for the connection $\ced B_i = 0$ to get 
\beq \label{eq:psiEOM}
\ced \psi_{ij} \we F^j = 0 \, ,
\eeq
where we used the Bianchi identity $\ced F^i \equiv 0$. In turn, substituting the solution for $B_i$ into the equation of $\psi^{ij}$ \eqref{eq:EOMpsi}, and multiplying on both sides with $\psi^{ij}$, we get  
\beq \label{eq:EOMpsiofF}
\frac{\iu}{2}\, F^i \we F^j = \ph \psi^{ik} \psi^{jl} \cH_{kl} \, .
\eeq 
Thus, after eliminating $B^i$ through its equation of motion, the full set of equations is 
\beq \label{eq:EOMtarget}
\ced \psi_{ij} \we F^j = 0 \, , \hspace{1cm} \frac{\iu}{2}\, F^i \we F^j = \ph \psi^{ik} \psi^{jl} \cH_{kl} \, , \hspace{1cm} \cH = 0  \, .
\eeq
Now the set \eqref{eq:EOMtarget} can be derived directly from the ``pure connection'' action
\beq \label{eq:SF}
S = \int \[ \frac{1}{2\iu}\, \psi_{ij} F^i \we F^j - \ph \cH(\bm{\psi})  \] \, .
\eeq
The action \eqref{eq:SF} is useful in revealing the following freedom in the definition of the $\cH(\bm{\psi})$ function. If one redefines $\bm{\psi}$ as follows
\beq \label{eq:psiinvredef}
\psi_{ij} \to \psi_{ij} + c \de_{ij} \, , 
\eeq
or, equivalently in matrix notation for $\psi^{ij}$, 
\beq \label{eq:psiredef}
\bm{\psi} \to \[ \bm{1} + c \bm{\psi} \]^{-1} \bm{\psi} \, ,  
\eeq
where $c$ is a complex constant, then the result is a modified function
\beq \label{eq:Hofpsitrans}
\cH(\bm{\psi}) \to \cH \( \[ \bm{1} + c \bm{\psi} \]^{-1} \bm{\psi} \) \, ,
\eeq
and a topological term $c F^i \we F^i$ in \eqref{eq:SF}, the latter being classically irrelevant. Therefore, two $\cH(\bm{\psi})$ functions related by the transformation \eqref{eq:Hofpsitrans} lead to the same classical theory. 

Another manipulation that leads to a classically equivalent theory is
\beq \label{eq:equivopt2}
\cH(\bm{\psi}) \to \frac{f(\cH(\bm{\psi}))}{f'(\cH(\bm{\psi}))} \, ,
\eeq
for any monotonic function $f$ satisfying $f(0) = 0$ and $f'(0) \neq 0$. Indeed, this means that the constraint is still $\cH = 0$, while the corresponding equations of motion are the same because the derivative of the new function is still $\cH_{ij}$
\beq
\frac{\pa }{\pa \psi^{ij}} \[ \frac{f(\cH)}{f'(\cH)} \] \equiv \[ 1 - \frac{f(\cH) \, f''(\cH)}{\[ f'(\cH) \]^2} \] \cH_{ij} = \cH_{ij} \, . 
\eeq

\subsection{Self-dual gravity} 

To conclude this section, let us describe another special choice of function $\cH(\bm{\psi})$ which shares some important properties with GR. The significance of this theory will become apparent in the following sections. This theory, named ``Self-Dual Gravity'' (SDG), is given by
\beq \label{eq:HSD*}
\cH_{\rm SDG}(\bm{\psi}) = -\, \psi_{ij} \delta^{ij}\, , 
\eeq
i.e. the same as GR \eqref{eq:HGR}, but for the inverse matrix $\psi_{ij}$, and where the cosmological constant term has been removed using the shift freedom \eqref{eq:psiinvredef}. This theory has been studied in detail in \cite{Krasnov:2016emc}. Inserting \eqref{eq:HSD*} inside the action \eqref{eq:SF} and the corresponding equations \eqref{eq:EOMtarget}, we see that $\psi_{ij}$ becomes a Lagrange multiplier enforcing the constraint
\beq \label{eq:psiEOMBB*}
\frac{\iu}{2}\, F^i \we F^j = \ph \de^{ij} \, .
\eeq
We will analyse the Lorentzian version of this theory, i.e. with the reality conditions imposed, in Section \ref{sec:reality}, and show that, in spite of it having the same number of degrees of freedom as GR, it is not a physical theory. However, both the Riemannian and split signature versions of this theory are meaningful and describe ``gravitational instantons", together with a field describing particles of opposite helicity and satisfying a linear field equation in the instanton background. See  \cite{Krasnov:2016emc} for more details. 

\section{Canonical formulation}
\label{sec:canonical}

\subsection{Space-time decomposition}

Our starting point are the equations of motion in the form \eqref{eq:EOMtarget}. We foliate space-time $x^{\mu} \to \{ t, x^{\al} \}$, define the ``magnetic'' fields
\beq
B^{i\al} := \frac{1}{2} \, \vep^{\al\be\ga} F_{\be\ga}^i \, ,
\eeq
and trade $\psi^{ij}$ for what will be shown to be the ``electric'' fields
\beq \label{eq:Edef}
E_i^{\al} := \psi_{ij} B^{j\al} \, ,
\eeq
which are constrained because of the symmetry of $\psi^{ij}$
\beq \label{eq:Poyntingcomp}
\vep_{\al\be\ga} E_i^{\be} B^{i\ga} = 0 \, .
\eeq
Thus, from now we have
\beq \label{eq:psidef}
\psi^{ij} = E_{\al}^i B^{j\al} = E_{\al}^j B^{i\al} \, ,
\eeq
where $E_{\al}^i$ is the inverse matrix of $E_i^{\al}$ and the symmetry in $ij$ is due to the constraint \eqref{eq:Poyntingcomp}. We will use this symmetry and \eqref{eq:Poyntingcomp} abundantly in what follows. Finally, we also define 
\beq \label{eq:Kalbedef}
H_{\al\be} := \iu F^i_{t\al} B_{i\be}  \, , 
\eeq
where $B_{i\al}$ is the inverse matrix of $B^{i\al}$. Finally, what will turn out to be the 3-metric density $\ti{q}_{\al\be}$ and (densitized) lapse function $\ti{N}$ of the underlying space-time are given by
\beq \label{eq:tiqalbetiNdef}
\ti{q}_{\al\be} := \cH_{ij} E^i_{\al} E^j_{\be}  \, , \hspace{1cm} \ph \equiv \ti{N} \ed^4 x \, .
\eeq
Here we have a first instance where $\cH_{ij}$ appears as an internal metric that relates a densitized dreibein to a densitized 3-metric. For instance, the inverse matrix $E_{\al}^i$ is related to $E_i^{\al}$ through the two metrics
\beq
E_{\al}^i \equiv \cH^{ij} \ti{q}_{\al\be} E_j^{\be} \, .
\eeq
In terms of these new fields, the equations \eqref{eq:EOMtarget} read respectively\footnote{To obtain the first three equations contract the 3-form components of the first equation of \eqref{eq:EOMtarget} with $\vep^{\mu\nu\ro\si}$, and for the fourth equation contract the free $i$ indices in the second equation of \eqref{eq:EOMtarget} with $B_{i\al}$ or $\bar{B}_{i\al}$, depending on the representation.}
\bea
D_{\al} E_i^{\al} & = & 0 \, ,  \label{Gauss} \\
D_t \psi_{ij} & = & - \iu \vep^{\al\be\ga} H_{\al\de} B^{k\de} B_{j\be} D_{\ga} \psi_{ik} \, , \label{eq:dtpsiinv*} \\
{\rm or}\,\,\, D_t \psi^{ij} & = & - \iu \vep^{\al\be\ga} H_{\al\de} E_k^{\de} E_{\be}^j D_{\ga} \psi^{ik} \, , \label{eq:dtpsi*} \\
H_{(\al\be)} & = & \ti{N} \ti{q}_{\al\be} \, , \label{eq:Ksym*} \\
\cH(\bm{\psi}(E,B)) & = & 0 \, ,  \label{eq:Hcomp} 
\eea
respectively, where
\beq
D_{\mu} X^i := \pa_{\mu} X^i + \vep^{i}{}_{jk} A^j_{\mu} X^k \, ,
\eeq
is the covariant derivative under local SO$(3,\Cs)$ transformations and we have used the Bianchi identity $D_{\al} B^{i\al} \equiv 0$.

We now introduce the notation 
\beq
N^{\al} := - \frac{\iu}{2} \, \vep^{\al\be\ga} H_{\be\ga} \, .
\eeq
Then \eqref{eq:Ksym*} gives
\beq \label{eq:Kalbe*}
H_{\al\be} = \ti{N} \ti{q}_{\al\be} + \iu \vep_{\al\be\ga} N^{\ga} \, .
\eeq
The underlying 4-metric of the theory is the Urbantke metric
\beq \label{eq:Urbantke}
\sqrt{-g} g_{\mu\nu} \propto \vep^{\ro\si\ka\la} \vep^{ijk} B_{i\mu\ro} B_{j\nu\si} B_{k\ka\la} \, , \hspace{1cm} \sqrt{-g} = \frac{\iu}{24}\,  \vep^{\mu\nu\ro\si} \cH^{ij} B_{i\mu\nu} B_{j\ro\si} \, ,
\eeq
which interprets $B_i$ as a ``cubic root'' of the metric. Our choice of the volume element containing $\cH^{ij}$ will be motivated later. This metric can be written in terms of the fields fields $\ti{N}$, $N^{\al}$ and $\ti{q}_{\al\be}$. Indeed, expressing the 4-metric \eqref{eq:Urbantke} in terms of $F^i$ through \eqref{eq:Bisol} and then using the definitions of this section, one finds\footnote{This can be checked easily using the computation of appendix \ref{app:RCtometric}, where the definition of $E_i^{\al}$ and $H_{\al\be}$ are consistent with the ones in this section if we take into account \eqref{eq:Bisol} and \eqref{eq:Edef}.}
\beq \label{eq:UrbADM}
g = - N^2 \ed t^2 + q_{\al\be} \( \ed x^{\al} + N^{\al} \ed t \) \( \ed x^{\be} + N^{\be} \ed t \) \, ,
\eeq 
where we have defined the weight zero counterparts
\beq \label{eq:qNdef}
q_{\al\be} := q \ti{q}_{\al\be}  \, , \hspace{1cm} N := q^{-1} \ti{N} \, ,
\eeq
with $q$ being the volume density squared  
\beq \label{eq:dettiqdef}
q := \sqrt{\det \ti{q}^{\al\be}} \equiv \det q_{\al\be} \, .
\eeq
We therefore recognize a lapse function $N$, a shift vector $N^{\al}$ and a 3-metric $q_{\al\be}$. The reality conditions to be imposed in the Lorentzian signature case will in particular imply that all quantities $N$, $N^{\al}$ and $q_{\al\be}$ are real. We will deal with this in Section \ref{sec:reality}. 

Having now access to a 4-geometry that is foliated 3+1, we have the notion of the normal vector to the spatial surfaces 
\beq \label{eq:ndef}
n := N^{-1} \( \pa_t - N^{\al} \pa_{\al} \) \, , 
\eeq
so it will be convenient to trade $A_t^i$ for the normal component
\beq
\te^i := - N n^{\mu} A_{\mu}^i \equiv - A_t^i + N^{\al} A_{\al}^i \, .
\eeq

\subsection{Evolution equations}

Let us next make the evolution equations more transparent. The one for $A_{\al}^i$ is found by contracting \eqref{eq:Kalbe*} with $B^{i\be}$
\beq \label{eq:dtA}
\dot{A}_{\al}^i = - \iu \ti{N} \ti{q}_{\al\be} B^{i\be} + \Lie_N A_{\al}^i - \pa_{\al} \te^i  \, ,
\eeq
which, in turn, allows to compute the one for $B^{i\al}$
\beq \label{eq:dtB}
\dot{B}^{i\al} = - \iu \vep^{\al\be\ga} D_{\be} \[ \ti{N} \ti{q}_{\ga\de} B^{i\de} \] + \Lie_N B^{i\al} + \vep^{i}{}_{jk} \te^j B^{k\al}  \, ,
\eeq
where $\Lie_N$ is the Lie derivative with respect to $N^{\al}$ and we have used the Bianchi identity $D_{\al} B^{i\al} \equiv 0$. Finally, with \eqref{eq:dtB} and \eqref{eq:Kalbe*} we can turn \eqref{eq:dtpsiinv*}, once contracted with $B^{j\alpha}$, into an evolution equation for $E_i^{\al}$
\beq \label{eq:dtE}
\dot{E}_i^{\al} = - \iu \vep^{\al\be\ga} D_{\be} \[ \ti{N} \ti{q}_{\ga\de} E_i^{\de} \] + \Lie_N E_i^{\al} + \vep^k{}_{ij} \te^j E_k^{\al} \, ,
\eeq
where we used \eqref{Gauss}. This equation is formally the same as the one for the magnetic fields \eqref{eq:dtB}. Note how the information of the specific theory under consideration, i.e. the function $\cH(\bm{\psi})$, is solely contained in the 3-metric density $\ti{q}_{\al\be}$ for the time evolution equations. Observe also how $N^{\al}$ and $\te^i$ simply generate the infinitesimal spatial diffeomorphisms and local Lorentz transformations, respectively.

\subsection{Canonical action} \label{sec:canaction}

In the notations introduced, the pure connection action \eqref{eq:SF} takes the following Hamiltonian form
\beq \label{eq:Scancomp*}
S = \int \ed^4 x \[ \frac{1}{\iu} \( E_i^{\al} \dot{A}_{\al}^i - \te^i \cG_i - N^{\al} \cD_{\al} \) - \ti{N} \cH  \]  \, , 
\eeq
where
\bea
\cH & \equiv & \cH(\bm{\psi}(E,B))  \, , \\
\cD_{\al} & := & \vep_{\al\be\ga} E_i^{\be} B^{i\ga} - A_{\al}^i \cG_i \equiv E_i^{\be} \( \pa_{\al} A_{\be}^i - \pa_{\be} A_{\al}^i \) - A_{\al}^i \pa_{\be} E_i^{\be} \, , \label{eq:Diffcomp} \\
\cG_i & := & D_{\al} E_i^{\al} \, . \label{eq:Gausscomp} 
\eea
The Euler-Lagrange equations obtained by extermising this action are \eqref{eq:dtA}, \eqref{eq:dtE}.
We now recognize in \eqref{eq:Scancomp*} the (de-densitized) Hamiltonian constraint $\cH$, the spatial diffeomorphism constraint $\cD_{\al}$ and the Gauss constraint $\cG_i$ that one expects from the local symmetries of the theory.

\subsection{Spatial metric from the constraint algebra}

The canonical formulation of diffeomorphism-invariant theories provides one with an alternative, dynamical notion of 3-metric, namely, the one which closes the hypersurface deformation algebra. Defining the holomorphic Poisson bracket
\beq \label{eq:holPB}
\{ \Ord, \Ord' \}_h := \int \[ \frac{\de \Ord}{\de A_{\al}^i} \frac{\de \Ord'}{\de E_i^{\al}} - \frac{\de \Ord'}{\de A_{\al}^i} \frac{\de \Ord}{\de E_i^{\al}} \] \, , 
\eeq
and the smeared constraints
\beq
\ti{H}[\ti{N}] := \int \ed^3 x\, \ti{N} \cH \, , \hspace{1cm} D[N] := \int \ed^3 x \, N^{\al} \cD_{\al} \, , \hspace{1cm} G[\te] := \int \ed^3 x \, \te^i \cG_i \, ,
\eeq
we find
\beq
\{ \ti{H}[\ti{N}], \ti{H}[\ti{N}'] \}_h = D[N] + G[\te] \, ,
\eeq
where
\beq
N^{\al} = q^{-1} q^{\al\be} \( \ti{N} \pa_{\be} \ti{N}' - \ti{N}' \pa_{\be} \ti{N} \) \, , \hspace{1cm} \te^i = q^{-1} q^{\al\be} A_{\al}^i \( \ti{N} \pa_{\be} \ti{N}' - \ti{N}' \pa_{\be} \ti{N} \) \, .
\eeq
This is the known result \cite{Krasnov:2007cq} that the 3-metric part of the Urbantke metric coincides with the 3-metric that closes the hypersurface deformation algebra. 

\subsection{Canonical shift transformations}

We note that the freedom in redefining $\psi_{ij}$ as in \eqref{eq:psiinvredef}, at the canonical level corresponds to redefining the momenta
\beq \label{eq:Eredef}
E_i^{\al} \to E_i^{\al} + c \delta_{ij} B^{j\al} \, ,
\eeq
because of \eqref{eq:Edef}. One can indeed check that, in the canonical action \eqref{eq:Scancomp*}, the term $E_i^{\al} \dot{A}_{\al}^i$ varies by a total time derivative (of the Chern-Simons 3-form of $A_{\al}^i$), the constraints $\cD_{\al}$ and $\cG_i$ are invariant, while $\cH$ does vary, thus leading to classically equivalent theories.

\subsection{Evolution equations for the metric} 

Using the evolution equations, we can compute the time-derivatives of the 3-metric, i.e. the ``geometrodynamics'' of the theory. We find 
\beq \label{eq:qtidotgen*}
\dot{\ti{q}}^{\al\be} = \Lie_N \ti{q}^{\al\be} - 2 \iu \ti{N} \[ E_i^{(\al} \vep^{\be)\ga\de} D_{\ga} E_{\de}^i -  C^{ij,m}_{kl} E^{-1} E_i^{\al} E_j^{\be} E_m^{\ga} D_{\ga} \psi^{kl} \] \, ,
\eeq
where
\beq
C^{ij,m}_{kl} := \vep^{mpq} \cH^{ir} \cH^{js} \[ \cH_{p(r} \cH_{s)q,kl} + \frac{1}{2}\,\cH_{p(k} \cH_{l)q,rs}  \] \, , \hspace{1cm} \cH_{ij,kl} := \frac{\pa \cH_{ij}}{\pa \psi^{kl}} \equiv \cH_{kl,ij}  \, .
\eeq
This is the evolution equation for the metric for a general modified theory. To interpret what is happening we now restrict our attention to a simpler four-parameter family of theories, which are completely tractable but general enough to illustrate the effects of the modification. 

\section{Minimal polynomial theories} 
\label{sec:polynomial}

\subsection{Parametrisation of $\cH(\psi)$}

Since $\cH(\bm{\psi})$ is an invariant matrix function, it can be expressed as a function of the scalars $\Tr \( \bm{\psi}^n \)$. However, thanks to the Cayley-Hamilton theorem for 3d matrices (or characteristic equation), $\bm{\psi}^{n>3}$ can be expressed in terms of $\bm{\psi}^{n\leq 3}$, so that we are left with only three independent combinations. We can therefore write, in full generality,
\beq \label{eq:HofpsiESB}
\cH(\bm{\psi}) \equiv f\( \psi_1, \psi_2, \psi_3 \) \, ,
\eeq
where
\bea
\psi_1 & := & \frac{1}{2}\,\vep_{ijk} \vep_{lmn} \de^{il} \de^{jm} \psi^{kn} \equiv \Tr \, \bm{\psi} \, , \\
\psi_2 & := & \frac{1}{2}\,\vep_{ijk} \vep_{lmn} \de^{il} \psi^{jm} \psi^{kn} \equiv \det \( \bm{\psi} \) \Tr \( \bm{\psi}^{-1} \) \, , \\
\psi_3 & := & \frac{1}{6}\,\vep_{ijk} \vep_{lmn} \psi^{il} \psi^{jm} \psi^{kn} \equiv \det \( \bm{\psi} \) \, ,
\eea
are the elementary symmetric polynomials of the three eigenvalues of $\bm{\psi}$. In terms of $E_i^{\al}$ and $B_i^{\al}$ we then find
\beq \label{eq:psi123ofpara}
\psi_1 := \frac{\( E,E,B \)}{\( E,E,E \)} \, , \hspace{1cm} \psi_2 := \frac{\( E,B,B \)}{\( E,E,E \)} \, , \hspace{1cm} \psi_3 := \frac{\( B,B,B \)}{\( E,E,E \)} \, ,
\eeq
where we have defined the symmetric triplet matrix product
\beq
\( X, Y, Z \) := \frac{\vep_{\al\be\ga} \vep^{ijk} X_i^{\al} Y_j^{\be} Z_k^{\ga}}{n_X! n_Y! n_Z!} \, ,
\eeq
with $n_{\star}$ denoting the multiplicity with which $\star$ appears. Given \eqref{eq:HofpsiESB}, we have that the $\cH_{ij}$ matrix controlling the 3-metric density \eqref{eq:tiqalbetiNdef} is given by
\beq
\bm{\cH} = \( f_1 + f_2 \psi_1 + f_3 \psi_2 \) \bm{1} - \( f_2 + f_3 \psi_1 \) \bm{\psi} + f_3 \bm{\psi}^2 \, , \hspace{1cm} f_{\star} := \frac{\pa f}{\pa \psi_{\star}} \, .
\eeq
For the inverse 3-metric density we then need the inverse matrix $\cH^{ij}$, which can be expressed as 
\beq \label{eq:HinvEBS}
\bm{\cH}^{-1} \equiv f_E \bm{1} + f_S \bm{\psi} + f_B \bm{\psi}^2 \, ,
\eeq
where 
\bea
f_E & := & \( \det \bm{\cH} \)^{-1} \[ f_1^2 + f_2 \( f_1 \psi_1 + f_2 \psi_2 + f_3 \psi_3 \) \]  \, , \nn  \\
f_S & := & \( \det \bm{\cH} \)^{-1} \[ f_1 f_2 + f_3 \( f_1 \psi_1 + f_2 \psi_2 + f_3 \psi_3 \) \]  \, , \label{eq:fESB123}  \\
f_B & := & \( \det \bm{\cH} \)^{-1} \[ f_2^2 - f_1 f_3 \] \, , \nn
\eea
and we have made abundant use of the characteristic equation
\beq \label{eq:chareq}
\bm{\psi}^3 - \psi_1 \bm{\psi}^2 + \psi_2 \bm{\psi} - \psi_3 \bm{1} \equiv 0 \, . 
\eeq
The advantage of the form \eqref{eq:HinvEBS} is that now the inverse 3-metric density becomes a linear combination of the three ``elementary'' inverse 3-metric densities
\beq \label{eq:qalbesimp}
\ti{q}^{\al\be} = f_E \de^{ij} E_i^{\al} E_j^{\be} + f_S E_i^{\al} B^{i\be} + f_B \de_{ij} B^{i\al} B^{j\be}  \, ,
\eeq
where $E_i^{\al} B^{i\be}$ is also symmetric by virtue of the constraint \eqref{eq:Poyntingcomp}. In the case of GR \eqref{eq:HGR}
\beq \label{eq:GRpsi123}
\cH_{\rm GR} = \la + \psi_1 \, , \hspace{1cm} \ti{q}_{\rm GR}^{\al\be} = \de^{ij} E_i^{\al} E_j^{\be} \, .
\eeq

\subsection{Four-parameter family of theories}

We now observe that GR has a very elegant property: its canonical action can be made polynomial and of minimal order in the electric and magnetic fields. This is already the case for the constraints $\cD_{\al}$ and $\cG_i$ for all theories. To achieve this polynomial form, we just need to trade $\ti{N}$ and $\cH$ for
\beq \label{eq:polredef}
\hat{N} := q^{-1} \ti{N} \, , \hspace{1cm} \hat{\cH} := q \cH \, ,
\eeq
respectively, noting in particular that $q \equiv E$, so that the corresponding term in the action is $\ti{N} \cH \equiv \hat{N} \hat{\cH}$, where  
\beq \label{eq:HhGR}
\hat{\cH} \equiv \la \( E, E, E \) + \( E, E, B \)  \, ,
\eeq
i.e. a polynomial of the minimal possible order in $E_i^{\al}$ and $B^{i\al}$, given the available contraction patterns. Although one can always perform such a redefinition,  the non-trivial aspect here is that the corresponding $\hat{N}$ is also real. This will be further discussed in Section \ref{sec:reality}. 

We now define the four-parameter family of natural extensions of GR. Indeed, the Hamiltonian constraint \eqref{eq:HhGR} is begging for the consideration of the $\( E, B, B \)$ and $\( B, B, B \)$ terms, while the inverse 3-metric density \eqref{eq:GRpsi123} can generalize to include the $E_i^{\al} B^{i\be}$ and $\de_{ij} B^{i\al} B^{j\be}$ terms. To build these ``minimal polynomial'' theories we first define a ``pre-Hamiltonian'' function
\beq \label{eq:tcHext}
\ti{\cH} := \la_0 + \la_1 \psi_1 + \la_2 \psi_2 + \la_3 \psi_3 \, ,  
\eeq
where the $\la$'s are complex constants, and whose derivative is given by
\beq
\ti{\cH}_{ij} := \frac{\pa \ti{\cH}}{\pa \psi^{ij}} = \( \la_1 + \la_2 \psi_1 + \la_3 \psi_2 \) \bm{1} - \( \la_2 + \la_3 \psi_1 \) \bm{\psi} + \la_3 \bm{\psi}^2 \, .
\eeq
We then consider the Hamiltonian function
\beq \label{eq:cHoftcH}
\cH := \frac{\ti{\cH}}{\det \ti{\bm{\cH}}} \, ,
\eeq
whose derivative is 
\beq
\bm{\cH} = \frac{\ti{\bm{\cH}}}{\det \ti{\bm{\cH}}} \, ,
\eeq
as we made use of the Hamiltonian constraint $\cH = 0$. We remind the reader that the boldface $\bm{\cH}$ stands for the matrix of first derivatives of the function $\cH(\psi)$. As a result, using \eqref{eq:HinvEBS}, \eqref{eq:fESB123} and again $\cH = 0$, the inverse matrix reads
\beq \label{eq:Hinvext}
\bm{\cH}^{-1} = (\det \ti{\bm{\cH}}) \, \ti{\bm{\cH}}^{-1} = \la_E \bm{1} + \la_S \bm{\psi} + \la_B \bm{\psi}^2 \, ,
\eeq
where
\beq \label{eq:laESB0123}
\la_E := \la_1^2 - \la_0 \la_2 \, , \hspace{1cm} \la_S := \la_1 \la_2 - \la_0 \la_3 \, , \hspace{1cm} \la_B := \la_2^2 - \la_1 \la_3 \, , 
\eeq
and thus leads to the most general minimal polynomial inverse 3-metric
\beq \label{eq:MPtiq}
\ti{q}^{\al\be} = \la_E \de^{ij} E_i^{\al} E_j^{\be} + \la_S E_i^{\al} B^{i\be} + \la_B \de_{ij} B^{i\al} B^{j\be} \, .
\eeq
Here we can also note that, using again $\cH = 0$ and \eqref{eq:chareq}, the determinant of $\ti{\bm{\cH}}$ involved in \eqref{eq:cHoftcH} can be expressed compactly as
\bea
\det \ti{\bm{\cH}} & = & \la_E \( \la_1 + \la_3 \psi_2 \) + \la_S \( \la_1 \psi_1 + \la_3 \psi_3 \) - \la_B \( \la_0 \psi_1 + \la_2 \psi_3 \) \nn \\
 & = & \la_E \( \la_1 + \la_3 \psi_2 \) - \la_S \( \la_0 + \la_2 \psi_2 \) - \la_B \( \la_0 \psi_1 + \la_2 \psi_3 \) \, .  \label{eq:dettcH}
\eea
As in the GR example shown above, we can next trade $\ti{N}$ and $\cH$ for the quantities given by \eqref{eq:polredef}, where now
\beq \label{eq:qofE}
q \equiv (\det \bm{\cH})^{-1/2} E = (\det \ti{\bm{\cH}})\, E \, ,
\eeq
so that $\hat{N}$ is also real and
\beq \label{eq:MPhatH}
\hat{\cH} = \la_0 \( E, E, E \) + \la_1 \( E, E, B \) + \la_2 \( E, B, B \) + \la_3 \( B, B, B \) \, ,
\eeq
is the most general minimal polynomial Hamiltonian. In particular, we see that GR, which corresponds to the case $\la_0 \in \Rs$, $\la_1 = 1$ and $\la_{2,3} = 0$, is actually the only theory with $\ti{q}^{\al\be} = \de^{ij} E_i^{\al} E_j^{\be}$, since $\la_S, \la_B = 0$ in \eqref{eq:laESB0123} forces $\la_2, \la_3 = 0$. 

\subsection{Classification of modifications}

Now remember that we still have the freedom to redefine the momenta \eqref{eq:Eredef} without altering the classical theory, because this transformation preserves the minimal polynomiality of the action.\footnote{Note that this transformation is less simple at the level of the $\cH(\bm{\psi})$ function, because $\bm{\psi}$ transforms in a more complicated way \eqref{eq:psiredef}.} The available theories are therefore separated into equivalence classes, where two theories are deemed equivalent if their actions can be related by \eqref{eq:Eredef}. Note also that this is the only redundancy around, because other options for generating equivalent theories, such as \eqref{eq:equivopt2}, are not available here, as they would spoil polynomiality of \eqref{eq:MPhatH}. Now the momentum shift \eqref{eq:Eredef} induces the following transformation in parameter space  
\bea
\la_0 & \to & \la_0 \, , \nn \\
\la_1 & \to & \la_1 + c \la_0 \, , \nn \\
\la_2 & \to & \la_2 + 2 c \la_1 + c^2 \la_2 \, , \label{eq:la0123map} \\
\la_3 & \to & \la_3 + 3 c \la_2 + 3 c^2 \la_1 + c^3 \la_0 \, , \nn
\eea
and
\bea
\la_E & \to & \la_E \, , \nn \\
\la_S & \to & \la_S + 2 c \la_E \, , \label{eq:laESBmap} \\
\la_B & \to & \la_B + c \la_S + c^2 \la_E \, . \nn
\eea
An important invariant combination under the described shifts is the ``discriminant'' of the quadratic expression \eqref{eq:Hinvext}
\beq \label{eq:discriminant}
\De := \la_S^2 - 4 \la_E \la_B \equiv \la_0^2 \la_3^2 - 3 \la_1^2 \la_2^2 + 4 \( \la_0 \la_2^3 + \la_3 \la_1^3 \) - 6 \la_0 \la_1 \la_2 \la_3 \, .
\eeq
The discriminant measures whether $\ti{q}^{\al\be}$ in \eqref{eq:MPtiq} is a perfect square. We will later see that the $\De=0$ case describes theories that are "metric" in the sense that their dynamics can be written solely in terms of the 3-metric and its time derivative (as well as the lapse and shift functions). The case $\De\not=0$ is that of essentially modified theories. 

Now the equivalence relations allows us to categorize the set of theories as follows. Noting that $\la_E$ is invariant under \eqref{eq:laESBmap}, we can split the theories into those with $\la_E \neq 0$, which are continuous deformations of GR, and those with $\la_E = 0$, which are not connected to that theory.
 
\noindent{\bf The case $\lambda_E\not=0$.}  In this case we can use the redefinition freedom to set $\la_B = 0$. This implies $\lambda_2^2=\lambda_1 \lambda_3$, which then gives $\lambda_1=\tau, \lambda_2=\si\tau^2, \lambda_3=\si^2\tau^3$, and therefore the Hamiltonian can be written as
\beq
\tilde{\cH}=\lambda_0 - \si^{-1} \[ 1 - \det(1+\si\tau \psi) \] \, .
\eeq
GR corresponds to the case $\si = 0$. If $\si \neq 0$, then we can set $\si = 1$ without loss of generality, by redefining $\la_0$ and $\ta$. This theory has been shown to arise as the result of the dimensional reduction on $S^3$ from a theory of 3-forms in seven dimensions, see \cite{Krasnov:2016wvc}. 

\noindent{\bf The case $\lambda_E=0$.}
This case can then be split further into those satisfying $\la_S \neq 0$ and those with $\la_S = 0$, which are invariant conditions once we have set $\la_E = 0$. In the case $\la_E, \la_S \not= 0$ we can use the transformations \eqref{eq:laESBmap} to set $\la_B=0$. Both $\la_E=0$ and $\la_B=0$ are achieved for $\la_1=\la_2=0$, which corresponds to the Hamiltonian
\beq
\tilde{\cH}=\lambda_0  + \la_3 \det(\psi) \, .
\eeq
This is the theory that was seen to arise \cite{Herfray:2016azk} as the dimensional reduction on $\R^3$ of a theory of 3-forms in seven dimensions. 

The case $\la_E, \la_S = 0$ is achieved by $\la_0=\la_1=0$. We can then use the momentum redefinition freedom to set $\la_3 = 0$. The result is the single theory
\beq \label{eq:BBHq}
\hat{\cH} = - \( E, B, B \) \, , \hspace{1cm} \ti{q}^{\al\be} = \de_{ij} B^{i\al} B^{j\al} \, , 
\eeq
where we have fixed the overall normalization such that $\la_B = 1$. Using \eqref{eq:cHoftcH} and \eqref{eq:dettcH}, this corresponds to
\beq \label{eq:dualGR}
\cH(\bm{\psi}) = - \frac{\psi_2}{\psi_3} \equiv  -\, \Tr \, (\bm{\psi}^{-1}) \, ,
\eeq
which is Self-Dual Gravity \eqref{eq:HSD*}. Comparing with \eqref{eq:HhGR}, we see that the equations of motion of this theory are formally the same as in GR with $\la = 0$, but with the electric and magnetic fields exchanged. 

In conclusion, the landscape of minimal polynomial theories, modulo the equivalence relation, is made of three disconnected parts: a stratum containing GR $\la_E \neq 0$, a 1-parameter continuum $\la_E = 0, \la_S \neq 0$ and one isolated point that is self-dual gravity (SDG) $\la_E, \la_S = 0$. Note also that the only theories with $\De = 0$ are GR and SDG. Indeed, $\De = 0$ if and only if $\ti{q}^{\al\be}$ is a complete square, so either $\ti{q}^{\al\be}$ can be brought to $\de^{ij} E_i^{\al} E_j^{\be}$ through a momentum shift (GR), or $\ti{q}^{\al\be} = \de_{ij} B^{i\al} B^{j\be}$ (SDG).

\subsection{Change of parametrisation}

We now trade the $\{ \la_0, \la_1, \la_2, \la_3 \}$ constants of the action for the set $\{ \mu,\ti{\mu},\nu,\ti{\nu} \}$ defined through
\beq \label{eq:latomunu}
\la_0 \equiv \frac{\mu^3 - \ti{\mu}^3}{\mu\nu - \ti{\mu} \ti{\nu}} \, , \hspace{0.5cm} \la_1 \equiv \frac{\mu^2 \ti{\nu} - \ti{\mu}^2 \nu}{\mu\nu - \ti{\mu} \ti{\nu}} \, , \hspace{0.5cm} \la_2 \equiv \frac{\mu \ti{\nu}^2 - \ti{\mu} \nu^2}{\mu\nu - \ti{\mu} \ti{\nu}} \, , \hspace{0.5cm} \la_3 \equiv \frac{\ti{\nu}^3 - \nu^3}{\mu\nu - \ti{\mu} \ti{\nu}} \, .
\eeq
For the 3-metric constants this implies 
\beq
\la_E \equiv \mu \ti{\mu} \, , \hspace{1cm} \la_S \equiv \mu \nu + \ti{\mu} \ti{\nu} \, , \hspace{1cm} \la_B \equiv \nu \ti{\nu} \, ,
\eeq
so that the discriminant \eqref{eq:discriminant} is given by
\beq
\De \equiv \( \mu \nu - \ti{\mu} \ti{\nu} \)^2 \, .
\eeq
Therefore, the substitution \eqref{eq:latomunu} is only valid for the theories with $\De \neq 0$, so it excludes a priori GR and Self-Dual Gravity. The 3-metric, however, is finite for all parameter values, and so will be all of the equations it satisfies, so in what follows we will be able to cover the $\De = 0$ cases as well. 

The advantage of this new parametrization is that the Hamiltonian \eqref{eq:MPhatH} and inverse 3-metric density \eqref{eq:MPtiq} can be expressed as
\beq \label{eq:hHqtgo}
\hat{\cH} \equiv \frac{\( \cE, \cE, \cE \) - \( \cB, \cB, \cB \)}{\sqrt{\De}} \, , \hspace{1cm} \ti{q}^{\al\be} \equiv \cE_i^{\al} \cB^{i\be} \, ,
\eeq
where
\beq
\cE_i^{\al} := \mu E_i^{\al} + \ti{\nu} \de_{ij} B^{j\al} \, , \hspace{1cm} \cB^{i\al} := \nu B^{i\al} + \ti{\mu} \de^{ij} E_j^{\al} \, .
\eeq
This structure will greatly simplify the computations to come. 

\subsection{Reformulation of evolution equations}

The general time evolution equations for the 3-metric \eqref{eq:qtidotgen*} are not transparent. In the case of minimal polynomial theories, however, these equations can be put in a form that readily admits a geometrodynamical interpretation, as we now show.

We first note that the determinant of the 3-metric, along with the Hamiltonian constraint \eqref{eq:hHqtgo}, leads to 
\beq \label{eq:detMdetq}
\cE = \cB = q \, ,
\eeq
where $\cE$ and $\cB$ denote the determinant of the respective matrices. Moreover, the Bianchi identity and the diffeomorphism and Gauss constraints imply
\beq
D_{\al} \cE_i^{\al} = D_{\al} \cB^{i\al} = 0 \, , \hspace{1cm} \vep_{\al\be\ga} \cE_i^{\be} \cB^{i\ga} = 0 \, , \hspace{1cm} \vep^i_{\,\,\,jk} \cE_{\al}^j \cB^{k\al} = 0 \, .
\eeq
In what follows we will always work at the level of the equations of motion, so the above constraints will always be understood and we will therefore not mention them explicitly every time. We next invoke the fully covariant derivative $\na_{\al}$, i.e. with respect to both local SO$(3,\C)$ transformations and spatial diffeomorphisms, that is compatible with the 3-metric $\na_{\ga} q_{\al\be} \equiv 0$, e.g.
\beq
\na_{\be} \cE_i^{\al} := D_{\be} \cE_i^{\al} + \Ga^{\al}_{\be\ga} \cE_i^{\ga} - \Ga^{\ga}_{\ga\be} \cE_i^{\al} \, ,
\eeq 
where $\Ga^{\al}_{\be\ga}$ are the Christoffel symbols of $q_{\al\be}$. We therefore have the useful property 
\beq \label{eq:MnaP}
\cE_i^{\al} \na_{\ga} \cB^{i\be} = - \cB^{i\be} \na_{\ga} \cE_i^{\al} \, ,
\eeq
and also the fact that all possible traces of this tensor density are zero
\bea
\cE_i^{\al} \na_{\be} \cB^{i\be} & = & 0 \, , \nn \\
\cE_i^{\al} \na_{\al} \cB^{i\be} & = & \na_{\al} \( \cE_i^{\al} \cB^{i\be} \) \propto \na_{\al} \ti{q}^{\al\be} \equiv 0 \, , \label{eq:traceless} \\
\ti{q}_{\al\be} \cE_i^{\al} \na_{\ga} \cB^{i\be} & \propto & \cB_{i\be} \na_{\ga} \cB^{i\be} \propto \na_{\ga} \cB = \na_{\ga} q \equiv 0 \, . \nn
\eea
We also introduce the volume 3-form
\beq
q_{\al\be\ga} := \sqrt{q} \vep_{\al\be\ga} \, ,
\eeq
and from now on the Greek indices are displaced using the 3-metric $q_{\al\be}$, except for $\cE_i^{\al}$ and $\cB^{i\al}$, since we will again use $\cE_{\al}^i$ and $\cB_{i\al}$ to denote their inverse matrices. In terms of $\cE_i^{\al}$, $\cB^{i\al}$ and $\na_{\al}$ the time-evolution equations \eqref{eq:dtA}, \eqref{eq:dtB} and \eqref{eq:dtE} read 
\bea 
\dot{A}_{\al}^i & = & - \frac{\iu N}{\sqrt{\De q}} \, q_{\al\be} \( \mu \cB^{i\be} - \ti{\mu} \de^{ij} \cE_j^{\be} \) + \Lie_N A_{\al}^i - \pa_{\al} \te^i  \, , \label{eq:dtcA} \\
\dot{\cB}^{i\al} & = & -\iu q^{\al\be}_{\,\,\,\,\,\,\,\ga} \na_{\be} \( N \cB^{i\ga} \) + \Lie_N \cB^{i\al} + \vep^i_{\,\,\,jk} \te^j \cB^{k\al} \, , \label{eq:dtcB} \\
\dot{\cE}_i^{\al} & = & -\iu q^{\al\be}_{\,\,\,\,\,\,\,\ga} \na_{\be} \( N \cE_i^{\ga} \) + \Lie_N \cE_i^{\al} + \vep^k_{ij} \te^j \cE_k^{\al} \, . \label{eq:dtcE}
\eea
where we remind that $N := \ti{N}/\sqrt{q} \equiv \sqrt{q} \hat{N}$. Next, we split the combination \eqref{eq:MnaP} into its symmetric and antisymmetric parts
\beq \label{eq:SKdef}
S_{\ga}^{\,\,\,\al\be} := \frac{1}{\sqrt{\De} q} \, \cE_i^{(\al} \na_{\ga} \cB^{i\be)} \, ,  \hspace{1cm}  K_{\al\be} := \frac{\iu}{2q} \, q_{\al\ga\de} \cE_i^{\ga} \na_{\be} \cB^{i\de} \, ,
\eeq
noting in particular that the latter is symmetric $\vep^{\al\be\ga} K_{\be\ga} = 0$, because of the tracelessness property \eqref{eq:traceless}. Also, notice that, in terms of $E_i^{\al}$ and $B^{i\al}$ 
\beq \label{eq:SofEB}
S_{\ga}^{\,\,\,\al\be} = \frac{1}{2q} \[ E_i^{\al} \na_{\ga} B^{i\be} - B^{i\al} \na_{\ga} E_i^{\be} \] = \frac{1}{2q} \[ E_i^{\al} D_{\ga} B^{i\be} - B^{i\al} D_{\ga} E_i^{\be} \] = \frac{1}{2q} \,E_i^{\al} E_j^{\be} D_{\ga} \psi^{ij} \, , 
\eeq
which shows that this tensor actually has a finite $\De \to 0$ limit. With \eqref{eq:SKdef} we can then express the spatial derivatives of $\cE_i^{\al}$ and $\cB^{i\al}$ as  
\beq \label{eq:naPnaMKS}
\na_{\ga} \cE_i^{\be} = - \[ \iu q_{\al}^{\,\,\,\be\de} K_{\ga\de} + \sqrt{\De} S_{\ga\al}^{\,\,\,\,\,\,\,\be} \] \cE_i^{\al}  \, , \hspace{1cm} \na_{\ga} \cB^{i\be} = - \[ \iu q_{\al}^{\,\,\,\be\de} K_{\ga\de} - \sqrt{\De} S_{\ga\al}^{\,\,\,\,\,\,\,\be} \] \cB^{i\al} \, .
\eeq
With this, inserting \eqref{eq:naPnaMKS} in the evolution equations \eqref{eq:dtcB} and \eqref{eq:dtcE} leads to
\bea 
\dot{\cE}_i^{\al} & = & N \( K \de^{\al}_{\be} - K^{\al}_{\be} + \sqrt{\De} S^{\al}_{\be} \) \cE_i^{\be} - \iu q^{\al\be}_{\,\,\,\,\,\,\,\ga} \cE_i^{\ga} \pa_{\be} N + \Lie_N \cE_i^{\al} + \vep^k_{\,\,\,ij} \te^j \cE_k^{\al} \, , \label{eq:dtcE2} \\
\dot{\cB}^{i\al} & = & N \( K \de^{\al}_{\be} - K^{\al}_{\be} - \sqrt{\De} S^{\al}_{\be} \) \cB^{i\be} - \iu q^{\al\be}_{\,\,\,\,\,\,\,\ga} \cB^{i\ga} \pa_{\be} N + \Lie_N \cB^{i\al} + \vep^i_{\,\,\,jk} \te^j \cB^{k\al} \, , \label{eq:dtcB2} 
\eea
where
\beq \label{eq:Si2def}
S_{\al\be} := \iu q_{\al\ga\de} S^{\ga\de}_{\,\,\,\,\,\,\be} \, ,
\eeq
is traceless $S_{\al}^{\al} \equiv 0$, because of the symmetry $S^{\al\be\ga} \equiv S^{\al\ga\be}$, and also symmetric $\vep^{\al\be\ga} S_{\be\ga} = 0$ because of \eqref{eq:traceless}

\subsection{Geometrodynamics}

We now have everything we need to compute the time evolution of the 3-metric. We first have the inverse 3-metric density \eqref{eq:hHqtgo}
\beq \label{eq:dotqtext}
\dot{\ti{q}}^{\al\be} = \Lie_N \ti{q}^{\al\be} + 2 N q \( q^{\al\be} K - K^{\al\be} \) \, .
\eeq
which in terms of the 3-metric reads
\beq \label{eq:dotqext}
\( \pa_t - \Lie_N \) q_{\al\be} = 2 N K_{\al\be} \, ,
\eeq
and reveals the geometric interpretation of $K_{\al\be}$ as the extrinsic curvature tensor. Using \eqref{eq:dtcA}, \eqref{eq:dtcB2}, \eqref{eq:dtcE2} and \eqref{eq:dotqext} we next find
\beq \label{eq:dotK0}
\( \pa_t - \Lie_N \) K_{\al\be} = \[ \na_{\al} \na_{\be} + K_{\al\ga} K_{\be}^{\ga} + 2 \De S_{[\al\ga]\de} S_{\be}^{\,\,\,\ga\de} + \iu q_{\al}^{\,\,\,\ga\de} \na_{\ga} K_{\de\be} - \frac{1}{2} \( V_{\al\be} - q_{\al\be} V \)  \] N \, ,
\eeq
where the curl of the extrinsic curvature arises from the time-derivative of the Christoffels in $\na$, and
\beq 
V^{\al\be} := - \frac{1}{\sqrt{\De} q} \[ \mu \de^{ij} \cE_i^{\al} \cE_j^{\be} - \ti{\mu} \de_{ij} \cB^{i\al} \cB^{j\be} \] \, , \hspace{1cm} V := V_{\al}^{\al} \, ,
\eeq
arises through the time-derivative of the spin connection in $\na$. Here too this quantity has a finite $\De \to 0$ limit, because in terms of $E_i^{\al}$ and $B_i^{\al}$
\beq \label{eq:VofEB}
q V^{\al\be} \equiv \la_0 E_i^{\al} E_i^{\be} + 2 \la_1 E_i^{\al} B_i^{\be} + \la_2 B_i^{\al} B_i^{\be} \, .
\eeq
Moreover, its trace is actually a 4-scalar (use \eqref{eq:psi123ofpara}, \eqref{eq:qofE} and \eqref{eq:dettcH})
\bea
V & = & - \frac{1}{\sqrt{\De} q} \[ \mu \( \cB, \cB, \cE \) - \ti{\mu} \( \cE, \cE, \cB \) \] \nn \\
 & = & - \frac{1}{q} \[ \la_E \( E, E, B \) + \la_S \( E, B, B \) + 3 \la_B \( B, B, B \) \] \nn \\
 & = & - \frac{\la_E \psi_1 + \la_S \psi_2 + 3 \la_B \psi_3}{\la_E \( \la_1 + \la_3 \psi_2 \) - \la_S \( \la_0 + \la_2 \psi_2 \) - \la_B \( \la_0 \psi_1 + \la_2 \psi_3 \)}  \, . \label{eq:TrVofEB}
\eea
We can then further develop the curl of $K_{\al\be}$
\bea
q_{\al}^{\,\,\,\ga\de} \na_{\ga} K_{\de\be} & \equiv & \frac{\iu}{2} \, \ti{q}_{\al\ga} \na_{\de} \( \cE_i^{\ga} \na_{\be} \cB_i^{\de} + \cB_i^{\ga} \na_{\be} \cE_i^{\de} \) \label{eq:Omcalc}  \\
 & = & \frac{\iu}{2} \, \ti{q}_{\al\ga} \( \na_{\de} \cE_i^{\ga} \na_{\be} \cB_i^{\de} + \na_{\de} \cB_i^{\ga} \na_{\be} \cE_i^{\de} + \cE_i^{\ga} \[ \na_{\de}, \na_{\be} \] \cB_i^{\de} + \cB_i^{\ga} \[ \na_{\de}, \na_{\be} \] \cE_i^{\de} \) \nn \\
 & = & \iu \[ K K_{\al\be} - K_{\al\ga} K_{\be}^{\ga} - \De S_{\ga\de\al} S_{\be}^{\,\,\,\ga\de} + R_{\al\be} - \frac{1}{2} \( V_{\al\be} + q_{\al\be} V \) \] \, , \nn 
\eea
where $R_{\al\be}$ is the Ricci tensor of $q_{\al\be}$. With this \eqref{eq:dotK0} becomes
\beq  \label{eq:dotK1}
\( \pa_t - \Lie_N \) K_{\al\be} - \[ 2 K_{\al\ga} K_{\be}^{\ga} - K K_{\al\be} - R_{\al\be} + \na_{\al} \na_{\be} \] N \approx N \[ \De S_{\al\ga\de} S_{\be}^{\,\,\,\ga\de} + q_{\al\be} V \] \, .
\eeq
The reader familiar with the 3+1 metric formulation of GR will recognize on the left-hand side the spatial part of the Ricci tensor of the 4-metric \eqref{eq:UrbADM}. We have therefore obtained the spatial part of the Einstein equation in the form $R_{\mu\nu} = T^{\rm eff}_{\mu\nu} - \frac{1}{2}\, g_{\mu\nu} T^{\rm eff}$, with an ``effective'' energy-momentum source $T_{\mu\nu}^{\rm eff}$ controlled by $V$ and $S_{\al\be\ga}$. In fact, one can also recover the time components, in the form $G_{\mu\nu} = T_{\mu\nu}$, by contracting \eqref{eq:Omcalc} with $q^{\al\be}$ and $q^{\al\be\ep}$. We thus obtain the full Einstein equations in ``standard ADM'' form \cite{York:1978gql} 
\beq \label{eq:modconstrGR}
\frac{1}{2} \[ R + K^2 - K_{\al\be} K^{\al\be} \] = \ro \, ,  \hspace{1cm} \na_{\al} K - \na_{\be} K_{\al}^{\be} = \cP_{\al} \, ,
\eeq
and 
\beq  \label{eq:dotK}
\( \pa_t - \Lie_N \) K_{\al\be} - \[ 2 K_{\al\ga} K_{\be}^{\ga} - K K_{\al\be} - R_{\al\be} + \na_{\al} \na_{\be} \] N = N \[ \cS_{\al\be} - \frac{1}{2}\, q_{\al\be} \( \cS - \ro \) \] \, ,
\eeq
where
\bea
\ro & := & n^{\mu} n^{\nu} T_{\mu\nu}^{\rm eff} \equiv \frac{\De}{2}\, S_{\al\be\ga} S^{\be\ga\al} + V \, , \label{eq:effE}  \\
\cP_{\al} & := & - n^{\mu} q_{\al}^{\nu} T_{\mu\nu}^{\rm eff} \equiv \iu \De q_{\al\be\ga} S_{\de\ep}^{\,\,\,\,\,\be} S^{\ga\de\ep} \, , \label{eq:effP} \\
\cS_{\al\be} & := & q_{\al}^{\mu} q_{\be}^{\nu} T_{\mu\nu}^{\rm eff} \equiv \De \( S_{\al\ga\de} S_{\be}^{\,\,\,\ga\de} - 2 S_{[\ga\de]\ep} S^{\ga\de\ep} \) - q_{\al\be} \ro  \, , \label{eq:effS} 
\eea
are the effective energy density $\ro$, momentum density $\cP_{\al}$ and stress density $\cS_{\al\be}$ in the $n^{\mu}$ frame \eqref{eq:ndef}, with normal projector $q_{\mu\nu} := g_{\mu\nu} + n_{\mu} n_{\nu}$. Using \eqref{eq:SofEB}, \eqref{eq:tiqalbetiNdef} and \eqref{eq:dtpsi} we then find the 4-covariant expression
\beq \label{eq:Teff}
T^{\rm eff}_{\mu\nu} = \frac{\De}{4} \[ \de_{\mu}^{\ro} \de_{\nu}^{\si} - \frac{1}{2}\, g_{\mu\nu} g^{\ro\si} \] \cH_{ik} \cH_{jl}  D_{\ro} \psi^{ij} D_{\si} \psi^{kl} - g_{\mu\nu} V \, .
\eeq
Remarkably, this takes the form of the energy-momentum tensor of a set of minimally coupled complex scalar fields $\psi^{ij}$, that are tensors in an internal space with metric $\cH_{ij}(\bm{\psi})$, and potential $V(\bm{\psi})$. Therefore, this expression can be formally obtained by varying the following ``matter'' action with respect to $g_{\mu\nu}$
\beq
S_m^{\rm eff} := \int \ed^4 x\, \sqrt{-g} \[ - \frac{1}{8}\, \De \, \cH_{ik}(\bm{\psi})\, \cH_{jl}(\bm{\psi})\, g^{\mu\nu} D_{\mu} \psi^{ij} D_{\nu} \psi^{kl} - V(\bm{\psi}) \] \, .
\eeq
This representation of the modified dynamics as that of GR with a non-trivial effective stress-energy tensor is one of the main results of this paper. It can can be compared to the results in \cite{Krasnov:2009ik} where it was explained how the modified theories can be similarly rewritten in explicitly metric terms. The results here are much more explicit, however. 

\section{Reality conditions}
\label{sec:reality}

We can now return to the second main objective of the present work, which is to analyse whether the condition that the 3-metric is real is an admissible condition for modified theories. This problem brings with itself many subtleties, which we must address beforehand. The first of such subtleties is how to deal with actions that involve complex fields in the case such actions are neither real nor holomorphic. Such actions arise when we add a real term imposing the reality conditions to the holomorphic terms describing the complexified modified theories.  

\subsection{Subtleties regarding complex non-holomorphic actions} \label{sec:actamb}

When manipulating actions involving a complex field $\ph$, the common and simplest approach is to consider $\ph$ and its conjugate $\bar{\ph}$ as independent variables, instead of the real and imaginary parts. At the level of the corresponding equations of motion, this is justified by the fact that the functional derivatives are simply related
\beq
\left. \frac{\de S}{\de \ph_{\rm Re}} \right|_{\ph_{\rm Im}} \equiv \left. \frac{\de S}{\de \ph} \right|_{\bar{\ph}} \left. \frac{\de \ph}{\de \ph_{\rm Re}} \right|_{\ph_{\rm Im}} + \left. \frac{\de S}{\de \bar{\ph}} \right|_{\ph} \left. \frac{\de \bar{\ph}}{\de \ph_{\rm Re}} \right|_{\ph_{\rm Im}} \equiv \left. \frac{\de S}{\de \ph} \right|_{\bar{\ph}} + \left. \frac{\de S}{\de \bar{\ph}} \right|_{\ph} \, ,
\eeq 
and, similarly,
\beq
\left. \frac{\de S}{\de \ph_{\rm Im}} \right|_{\ph_{\rm Re}} \equiv \left. \iu \[ \frac{\de S}{\de \ph} \right|_{\bar{\ph}} - \left. \frac{\de S}{\de \bar{\ph}} \right|_{\ph} \] \, ,
\eeq
so that setting either type of variation to zero leads to the same equations. Importantly, this does not require the action to be real, as is usually the case. If, however, the action is real, then we also have the relation
\beq
\frac{\de S}{\de \bar{\ph}} \equiv \overline{\frac{\de S}{\de \ph}} \, ,
\eeq
which implies that the equation $\de S/ \de \ph = 0$ is equivalent to $\de S / \de \bar{\ph} = 0$ and that there are therefore only two real equations of motion. Another special case is when the action is holomorphic
\beq
\frac{\de S}{\de \ph} \equiv \frac{\de S}{\de \ph_{\rm Re}} \equiv -\, \iu\, \frac{\de S}{\de \ph_{\rm Im}} \, , \hspace{1cm} \frac{\de S}{\de \bar{\ph}} \equiv 0 \, , 
\eeq
and so here too we have as many real equations of motion as the number of real fields. This is no longer true for complex non-holomorphic actions, which is the case of interest here. Indeed, since the action is not real we have that $\de S/ \de \ph_{\rm Re} \in \Cs$, and since the action is not holomorphic we have that $\de S/ \de \ph_{\rm Re} \,\, {\not \propto} \,\, \de S/ \de \ph_{\rm Im}$. Taking the real and imaginary parts of both $\de S/ \de \ph_{\rm Re}$ and $\de S/ \de \ph_{\rm Im}$ we obtain four real equations of motion for only two real fields $\ph_{\rm Re}$ and $\ph_{\rm Im}$. Therefore, such a system is generically over-determined, unless part of the equations are dependent. The presence of more equations is not surprising, since a complex non-holomorphic action is essentially two real actions. 

This mismatch between the number of fields and the number of equations makes the procedure of integrating fields in and out of such an action ambiguous. To see this, say we wish to integrate $\ph$ out. Assuming it enters the action algebraically and non-linearly, the solution to its equation of motion $\de S / \de \ph = 0$ allows us to eliminate it, thus also eliminating $\bar{\ph}$ through the conjugate of that solution. The problem is that the equation of motion of the latter $\de S / \de \bar{\ph} = 0$ is independent from $\de S / \de \ph = 0$ and is no longer taken into account, since $S$ is now independent of both $\ph$ and $\bar{\ph}$. We have therefore lost part of the equations of motion. Conversely, when integrating in a new field, such as in a Legendre transform of the action, it can happen that we end up with more equations than we had in the original theory. This is why in this paper we always work out the equations of motion, rather than perform manipulations with the Lagrangian. 

Let us now be more specific by focusing on the the subclass of actions we will consider in this paper. Our aim is to impose reality constraints on the solutions of a holomorphic theory and to do so through a variational principle. We therefore consider actions of the form 
\beq \label{eq:S}
S := S_h + S_c \, ,
\eeq
where $S_h \equiv S_h[\ph]$ is a holomorphic functional of the fields collectively denoted by $\ph$, i.e. $\de S/\de \bar{\ph} \equiv 0$, while $S_c$ is real and exclusively made of ``reality constraints''
\beq
S_c := \iu \int \ch \( f[\ph, \pa \ph] - \overline{f[\ph, \pa \ph] } \) \, ,
\eeq
where $\ch$ denotes real Lagrange multiplier fields imposing the reality constraints
\beq \label{eq:Imf0}
{\rm Im}\, f \(\ph, \pa \ph \) = 0 \, ,
\eeq
and $f$ is a holomorphic function of its variables. Varying with respect to $\ph$, $\bar{\ph}$ and $\ch$, the equations of motion of this action are respectively,
\beq \label{eq:EOMSex}
\frac{\de S_h}{\de \ph} + \iu \[ \ch\, \frac{\pa f}{\pa \ph} - \pa \cdot \( \ch\, \frac{\pa f}{\pa (\pa\ph)} \) \] = 0 \, , \hspace{1cm} \overline{\ch \, \frac{\pa f}{\pa \ph} - \pa \cdot \( \ch\, \frac{\pa f}{\pa (\pa\ph)} \)} = 0  \, , \hspace{1cm} {\rm Im}\, f \( \ph,\pa \ph \) = 0 \, .
\eeq
Inserting the conjugate of the second equation into the first equation we find the equivalent system
\beq \label{eq:EOMSex2}
\frac{\de S_h}{\de \ph} = 0 \, , \hspace{1cm} \ch \, \frac{\pa f}{\pa \ph} - \pa \cdot \( \ch\, \frac{\pa f}{\pa (\pa\ph)} \) = 0  \, , \hspace{1cm} {\rm Im}\, f \( \ph,\pa \ph \) = 0 \, ,
\eeq
which are the equations of motion of the holomorphic theory $S_h$ supplemented by the reality constraints \eqref{eq:Imf0}, as desired, plus some extra equations involving the Lagrange multipliers $\ch$. Note that it is the equations of motion of the conjugate fields $\bar{\ph}$ that precisely make the $\ch$-dependent part of the equations decouple. Importantly, this would not be the case if we had also included the anti-holomorphic sector $\bar{S}_h$, so this is really a specificity of ``holomorphic + reality constraint'' actions. 

We must still pay attention to the extra equations in \eqref{eq:EOMSex2} for the Lagrange multipliers $\ch$, which are not necessarily harmless. Indeed, if $\pa f/\pa (\pa\ph) \neq 0$, then part of these equations might include time-derivatives of $\ch$, so that these fields possibly contain degrees of freedom. These are not present in the holomorphic theory we were initially trying to constrain, so in that case the action \eqref{eq:S} would not be appropriate. If, however, the equation for $\ch$ contains no time derivatives of that field, then setting trivial boundary conditions implies $\ch = 0$ everywhere, because of the linearity of the equations. In fact, as we will see, for our defining action $\ch = 0$ will be the only possible solution. Therefore, the extra equations for $\ch$ will all be solved without introducing spurious degrees of freedom and without imposing any extra conditions on $\ph$. The result are the holomorphic theory equations for $\ph$ subject to the reality constraints through a variational principle, as wished. 

\subsection{$BF$ formulation with reality conditions}

We now supplement the holomorphic action \eqref{eq:Sh} with a real term that imposes the reality conditions. 
\beq \label{eq:Sc}
S_c := \int \[ \ch\, {\rm Re} \[ \cH^{ij} B_i \we B_j \] + \ch^{ij} B_i \we \bar{B}_j \] \, .
\eeq
where
$\ch \in \Rs$ and $\ch^{ij} \equiv \bar{\ch}^{ji}$ are 0-forms. A similar formulation of the reality conditions was considered in \cite{Lewandowski_2000}.
The fields $\ch, \ch^{ij}$ are the Lagrange multipliers imposing the scalar and tensor reality constraints 
\beq \label{eq:realconstr}
{\rm Re} \[ \cH^{ij} B_i \we B_j \] = 0 \, , \hspace{1cm} B_i \we \bar{B}_j = 0 \, ,
\eeq
respectively. Inserting now the holomorphic equation \eqref{eq:EOMpsi} in the scalar constraint we see that it simply amounts to the reality of $\ph$, so a simpler equivalent action would be 
\beq \label{eq:Ssimp}
S = \int \[ \frac{1}{\iu} \( B_i \we F^i - \frac{1}{2} \, \psi^{ij} B_i \we B_j \) - \ph \cH(\bm{\psi}) + \ch^{ij} B_i \we \bar{B}_j \] \, , \hspace{1cm} \ph \in \Rs \, ,
\eeq
since it leads to the same equations of motion. The disadvantage of this formulation, however, is that one can no longer see the ``holomorphic + reality constraint'' structure so easily. 

Let us now discuss in what sense \eqref{eq:realconstr} are the desired reality constraints. As we show in detail in appendix \ref{app:RCtometric}, \eqref{eq:realconstr} is equivalent to the much more transparent statement of the reality of the Urbantke metric \eqref{eq:Urbantke}
\beq
{\rm Im} \, g_{\mu\nu} = 0 \, .
\eeq
In particular, the scalar reality constraint makes the 4-volume density real, while the tensor constraint makes the conformal part of the metric real. 

Let us now show that the introduced Lagrange multipliers $\ch$ and $\ch^{ij}$ must be zero when all other field equations are satisfied. Thus, the introduction of these fields into the action does not add degrees of freedom, in agreement with the general discussion of the previous subsection. We take the equation of motion of $\bar{B}_i$
\beq \label{eq:EOMBbar}
\ch \cH^{ij} B_j + \ch^{ij} \bar{B}_j = 0 \, , 
\eeq
and first wedge it with $B_k$, using \eqref{eq:realconstr},
\beq
\ch \cH^{ij} B_j \we B_k = 0 \, ,
\eeq
which leads to $\ch = 0$, since $B_i \we B_j$ is invertible on the solutions of interest. Next, wedging \eqref{eq:EOMBbar} with $\bar{B}_k$ and using again \eqref{eq:realconstr} and $\ch = 0$, we find
\beq
\ch^{ij} \bar{B}_j \we \bar{B}_k = 0 \, ,
\eeq
which therefore implies $\ch^{ij} = 0$. 

We now make some general remarks. It is known that the reality conditions \eqref{eq:realconstr} are the correct Lorentzian signature reality conditions for the unmodified GR. When the theory is modified, it is far from clear whether there exist compatible reality constraints, i.e. constraints that select a non-trivial subset of the solutions of the holomorphic theory. As we already mentioned in the previous subsection, complex non-holomorphic actions will in general lead to over-determined equations, unless some of these equations are not independent. Secondly, even if compatible reality constraints exist, there is a priori no reason to expect them to be the same as the ones of GR. As we will see, however, the tensor constraint is simply too rigid to be deformed in a continuous way, while any generalization of the scalar constraint can be reabsorbed in a renormalization of the $\cH(\bm{\psi})$ function, so that it does not provide new directions in theory space. Therefore, one can choose the scalar constraint of most convenience, our choice \eqref{eq:realconstr} being the simplest one, as it leads to ${\rm Im}\, \ph = 0$. We postpone the demonstration of this claim to section \ref{sec:genRC}, because it will be a lot easier if we use elements that will be provided in the following subsection. 

Thus, while the reality conditions of the modified theories may be different from those of GR, the GR reality constraints in the form \eqref{eq:realconstr} are too rigid to admit a deformation. It is not excluded that there exists a different formulation of the GR reality constraints that does admit a deformation, but this is not known at present. The logic of what follows is to verify whether the reality conditions \eqref{eq:realconstr} are compatible with the dynamics of the modified theories. 

\subsection{Pure connection formulation}

Let us also discuss how the reality conditions can be stated in the pure connection formalism. We can substitute the solution for $B^i$ into the reality conditions \eqref{eq:realconstr} to find
\beq
{\rm Re} \[ \psi_{ik} \psi_{jl} \cH^{ij} F^k \we F^l \] = 0 \, , \hspace{1cm} F^i \we \bar{F}^j = 0 \, ,
\eeq
where we have multiplied the latter by $\psi^{ik}$ from the right and $\bar{\psi}^{jl}$ from the left. Finally, using \eqref{eq:EOMpsiofF} we can simplify the scalar constraint to
\beq \label{eq:Impheq}
{\rm Im} \, \ph = 0 \, .
\eeq
Thus, after eliminating $B_i$ through its equation of motion, the full set of equations of the theory with reality conditions imposed is $\ch, \ch^{ij} = 0$ and
\beq \label{eqs-reality}
\ced \psi_{ij} \we F^j = 0 \, , \hspace{0.5cm} \frac{\iu}{2}\, F^i \we F^j = \ph \psi^{ik} \psi^{jl} \cH_{kl} \, , \hspace{0.5cm} \cH = 0 \, ,  \hspace{0.5cm} {\rm Im} \, \ph = 0 \, , \hspace{0.5cm} F^i \we \bar{F}^j = 0 \, .
\eeq
Note that this set  can be derived directly from the ``pure connection'' action
\beq \label{eq:SF*}
S = \int \[ \frac{1}{2\iu}\, \psi_{ij} F^i \we F^j - \ph \cH(\bm{\psi}) + \ze_{ij} F^i \we \bar{F}^j \] \, ,  \hspace{1cm} \ph \in \Rs \, , 
\eeq
with $\ze_{ij} \equiv \bar{\ze}_{ji}$ being the Lagrange multilpier 0-form implementing the tensor reality constraint in this formulation. The only problem with this action is that the extra equation for $\ze_{ij}$, which comes from the variation with respect to $\bar{A}^i$, is a differential one
\beq
\ced \ze_{ij} \we \bar{F}^j = 0 \, ,
\eeq
since the tensor constraint is now a differential equation. This means that $\ze_{ij}$ contains degrees of freedom which were not present in the original theory. For this reason the action  is not equivalent to the original one and should only be seen as a compact way of describing the full set of equations. 

Finally, one more instructive aspect of the action \eqref{eq:SF*} is that it illustrates how reality constraints in complex non-holomorphic theories behave differently than usual constraints. We first observe that \eqref{eq:SF*} can be interpreted as the result of integrating out $B_i$ in \eqref{eq:Ssimp}, since that field appears quadratically and we have eliminated it through its own equation of motion \eqref{eq:Bisol}. Naively, one would expect such a manipulation to make the resulting action non-linear in the Lagrange multiplier $\ch^{ij}$, so that it no longer imposes a constraint and can further be integrated out. This is for instance what happens with $\psi^{ij}$ for the neighbors of complex GR, leading in particular to \eqref{eq:BFpV}, as we discussed in the previous subsection. Here, instead, we see that after integrating out $B_i$ the constraint imposed by $\ch^{ij}$ remains, even though it now affects different fields. This occurs precisely because $\ch^{ij}$ does not appear in the equation of motion of $B_i$, thanks to the equation of motion of $\bar{B}_i$ which preserves the holomorphic equations. One can therefore think of \eqref{eq:SF*} as the result of integrating out $B_i$ in \eqref{eq:Ssimp} using its holomorphic equation of motion, up to the subtlety that $\ze_{ij}$ cannot be interpreted as $\psi_{ik} \psi_{jl} \ch^{kl}$, because it satisfies a different equation with a different set of solutions. The important point here is that reality constraints in complex non-holomorphic theories survive the elimination of the field on which they act. 

\subsection{Rigidity of the reality conditions}  \label{sec:genRC}

We now dispose of the necessary tools to prove our claim that there is no room for generalization of the reality constraints \eqref{eq:realconstr}. We start with the scalar one, whose generalization from the GR case ${\rm Re} \[ \de^{ij} B_i \we B_j \] = 0$ is  
\beq \label{eq:genscalconstr}
{\rm Re} \[ C(\bm{Y}) \] = 0 \, ,
\eeq
where the $\bm{Y}$ matrix has been defined in \eqref{eq:BFpV}. Here $C$ is an invariant matrix function of $\bm{Y}$ that must also be homogeneous $C(\al \bm{Y}) = \al^d C(\bm{Y})$, for some degree $d$, for \eqref{eq:genscalconstr} to be invariant under diffeomorphisms. Without loss of generality, we can set $d = 1$ for definiteness, which in practice can be achieved by simply replacing $C \to C^{1/d}$ in \eqref{eq:genscalconstr}. The reality constraint part of the action then reads
\beq \label{eq:Scgen}
S_c = \int \[ \ch\, {\rm Re} \[ C(\bm{Y}) \] + \ch^{ij} B_i \we \bar{B}_j \] \, .
\eeq  
Note, in particular, that we still have $\ch, \ch^{ij} = 0$ as the only solutions, because \eqref{eq:EOMBbar} is modified to $\ch C^{ij} B_j + \ch^{ij} \bar{B}_j = 0$, where $C^{ij} := \pa C/ \pa Y_{ij}$ is invertible. We then realize that we can express the same constraint, while remaining quadratic in $B_i$, at the price of introducing a $\bm{\psi}$ dependence
\beq \label{eq:CYtoSBB}
C(\bm{Y}) \to S^{ij}(\bm{\psi})\, B_i \we B_j \, ,
\eeq
where $S^{ij}$ is a covariant matrix function of $\bm{\psi}$. Indeed, if one writes down the equations of motion in both cases and eliminates $\bm{\psi}$ using its own equation of motion $\psi^{ij} = \psi^{ij}(\bm{Y}/\ph)$, and then similarly for $\ph$, then the action takes the form \eqref{eq:BFpV} plus \eqref{eq:Scgen} in both cases. Notice that our choice in \eqref{eq:realconstr} is now a member of the class \eqref{eq:CYtoSBB}.

Now that we brought the most general constraint to the form \eqref{eq:CYtoSBB}, we can repeat the exercise of the previous subsection of eliminating $B_i$ in the equations of motion. The only difference with respect to \eqref{eq:EOMtarget} lies in the constraint on $\ph$, which now reads
\beq \label{eq:Impheq2}
{\rm Im} \[ \ph \cH_{ij} S^{ij} \] = 0 \, ,
\eeq
instead of \eqref{eq:Impheq}. However, we can then trade $\ph$ for $\ti{\ph} = \ph \cH_{ij} S^{ij}$, so that the equations take the form 
\beq \label{eq:EOMtarget2}
\ced \psi_{ij} \we F^j = 0 \, , \hspace{0.5cm} \frac{\iu}{2}\, F^i \we F^j = \ti{\ph} \psi^{ik} \psi^{jl} \ti{\cH}_{kl} \, , \hspace{0.5cm} \ti{\cH} = 0 \, ,  \hspace{0.5cm} {\rm Im} \, \ti{\ph} = 0 \, , \hspace{0.5cm} F^i \we \bar{F}^j = 0 \, ,
\eeq
where we have defined the new matrix functions of $\bm{\psi}$
\beq
\ti{\cH} := \frac{\cH}{\cH_{ij} S^{ij}} \, , \hspace{1cm} \ti{\cH}_{ij} := \frac{\pa \ti{\cH}}{\pa \psi^{ij}} \equiv \frac{1}{\cH_{kl} S^{kl}} \, \frac{\pa \cH}{\pa \psi^{ij}} - \frac{\cH}{\( \cH_{kl} S^{kl} \)^2} \, \frac{\pa \( \cH_{mn} S^{mn} \)}{\pa \psi^{ij}} = \frac{\cH_{ij}}{\cH_{kl} S^{kl}}    \, .
\eeq
Note that we have used the fact that $\cH_{ij} S^{ij} \neq 0$ on the solutions, so that the constraint actually remains $\cH = 0$. Thus, the equations take the exact same form as for our choice $S^{ij} = \cH^{ij}$, with the deviation $\cH_{ij} S^{ij}$ being completely reabsorbed in a renormalization of $\cH$. This proves that generalizing the scalar reality constraint does not lead to new directions in theory space, i.e. beyond those one can explore with $\cH$. 

Let us next now consider the tensor constraint. Following the same logic as for the scalar one, we are looking for generalizations involving the $\bm{\psi}$ matrix. This is hard to do in the form $B_i \we \bar{B}_j = 0$, since it couples the two representations. In particular, it is not possible to write down a continuous modification of that equation that will not also admit $B_i \we \bar{B}_j = 0$ as a solution. We therefore turn our attention to its alternative formulation as the reality of the Urbantke comformal class of metrics \eqref{eq:Urbantke}. The analogous generalization of \eqref{eq:CYtoSBB} is then
\beq \label{eq:ghatT}
{\rm Im} \[ C\, \vep^{\ro\si\ka\la}\, T^{ijk}(\bm{\psi}) \, B_{i\mu\ro} B_{j\nu\si} B_{k\ka\la} \] = 0 \, ,
\eeq
for some complex comformal factor $C$. Note that here $T^{ijk} \equiv - T^{jik}$ for the metric to be symmetric, which means that we can express it equivalently as
\beq \label{eq:TofT}
T^{ijk} \equiv \vep^{ijl} \de_{lm} T^{mk} \, , \hspace{1cm} T^{ij} := \frac{1}{2}\, \vep^i_{\,\,\,kl} T^{klj} \, ,
\eeq
where now $T^{ij} \equiv T^{ji}$ since it is a covariant matrix function of $\psi^{ij}$. But then, repeating the computation of appendix \ref{app:RCtometric} for \eqref{eq:ghatT}, we notice that it simply leads to an extra overall factor $\sim \de_{ij} T^{ij}$, so that the conformal class of metrics is exactly the same. Thus, the tensor constraint is not modifiable. There is an underlying geometrical reason behind this rigidity. Indeed, it is well-known that the conformal class of the Urbantke metric \eqref{eq:Urbantke} is the unique class with respect to which the $B_i$ 2-forms are self-dual 
\beq \label{eq:gsdB}
\frac{1}{2} \, \vep^{\mu\nu\ro\si} B_{i\ro\si} \equiv \iu \sqrt{-g} g^{\mu\ro} g^{\nu\si} B_{i\ro\si} \, .
\eeq
This is indeed a condition on the conformal class only, since it is invariant under a conformal transformation of the metric. It constitutes an equivalent, purely geometric definition of that conformal class, whose uniqueness explains why it is impossible to deform that 4-metric.

\subsection{Constrained Self-Dual Gravity} \label{sec:CSDgravity}

Before we address the question of compatibility of the reality conditions and the modified dynamics, let us analyse the effect of reality conditions in the case of Self-Dual Gravity (SDG). We recall that this is the theory given by
\beq \label{eq:HSD}
\cH_{\rm SDG}(\bm{\psi}) = -\, \psi_{ij} \delta^{ij}\, , 
\eeq
i.e. the same as GR \eqref{eq:HGR}, but for the inverse matrix $\psi_{ij}$, and where the cosmological constant term has been removed using the shift freedom \eqref{eq:psiinvredef}. The holomorphic version of this theory has already been studied in detail in \cite{Krasnov:2016emc}. We will call its version with reality conditions implemented ``constrained Self-Dual Gravity'' (CSDG). Using \eqref{eq:HSD} in the action \eqref{eq:SF} and the corresponding equations \eqref{eq:EOMtarget}, we see that $\psi_{ij}$ becomes a Lagrange multiplier enforcing the constraint
\beq \label{eq:psiEOMBB}
\frac{\iu}{2}\, F^i \we F^j = \ph \de^{ij} \, .
\eeq
Combined with the reality constraint $F^i \we \bar{F}^j = 0$, these equations can be compactly expressed in terms of the real spin connection curvature \eqref{eq:FIJ}
\beq \label{eq:HGRdualconstrto}
F^{IJ} \we F^{KL} = - \,\vep^{IJKL} \ph   \, .
\eeq
The reader familiar with the real $BF$ formulation of GR \cite{Plebanski:1977zz,DePietri:1998hnx} will then immediately notice that \eqref{eq:HGRdualconstrto} takes the form of the simplicity constraint, but now for the $F$-field instead of the $B$-field. The solution is therefore
\beq \label{eq:FBBsol1}
F^{IJ} = C\, e^I \we e^J \, , \hspace{1cm}   \ph = \frac{C^2}{24} \, \vep_{IJKL}\, e^I \we e^J \we e^K \we e^L \, ,
\eeq
or
\beq \label{eq:FBBsol2}
F^{IJ} = \frac{C}{2}\, \vep^{IJ}_{\,\,\,\,\,\,KL}\, e^K \we e^L \, , \hspace{1cm}   \ph = -\frac{C^2}{24} \, \vep_{IJKL}\, e^I \we e^J \we e^K \we e^L \, ,
\eeq
for a set of real vierbein 1-forms $e^I$ and a constant $C$. Taking the covariant exterior derivative of the first equation in \eqref{eq:FBBsol1} and \eqref{eq:FBBsol2} we find, after some algebraic manipulations, $\ced e^I = 0$, meaning that $A^{IJ}$ is the torsion-free spin connection of the vierbein $e^I$. On the other hand, wedging the first equation of \eqref{eq:FBBsol1} and \eqref{eq:FBBsol2} with $e_J$ and using $0 = \ced^2 e^I \equiv F^I_{\,\,\,J} \we e^J$, we find that the only non-trivial possibility is \eqref{eq:FBBsol1}. This is now the defining equation of a maximally symmetric space-time with the scalar curvature $C$. Thus, in stark contrast with GR, in CSDG there are no degrees of freedom in the spin connection $A^i$. Instead, these lie in the $\psi_{ij}$ field which satisfies a first-order linear evolution equation (the first of \eqref{eq:EOMtarget}) on an (A)dS background, so this is a linear theory in disguise, even though its action is invariant under the full local symmetries. Having a linear theory is not a problem, since one could in principle include interactions with matter, as in the case of electrodynamics for instance. However, the problem here is that the action is linear in the only dynamical field $\psi_{ij}$, meaning that the corresponding energy of the fluctuations cannot be bounded from below. Therefore CSDG is not physical.

\subsection{Canonical formalism and reality conditions}

We now address the main question, which is to analyse the dynamics of the modified theories with reality conditions imposed. The full set of field equations in the canonical formalism is now
\bea
D_{\al} E_i^{\al} & = & 0 \, ,  \\
D_t \psi_{ij} & = & - \iu \vep^{\al\be\ga} H_{\al\de} B^{k\de} B_{j\be} D_{\ga} \psi_{ik} \, , \label{eq:dtpsiinv} \\
{\rm or}\,\,\, D_t \psi^{ij} & = & - \iu \vep^{\al\be\ga} H_{\al\de} E_k^{\de} E_{\be}^j D_{\ga} \psi^{ik} \, , \label{eq:dtpsi} \\
H_{(\al\be)} & = & \ti{N} \ti{q}_{\al\be} \, , \label{eq:Ksym} \\
\cH(\bm{\psi}(E,B)) & = & 0 \, ,  \label{eq:Hcomp} \\
{\rm Im} \, \ti{N} & = & 0 \, , \\
H_{\al\be} & = & \bar{H}_{\be\al}  \, , \label{eq:Kherm}
\eea
Note that the tensor reality constraint, that is the last equation of \eqref{eqs-reality}, becomes the hermiticity condition  \eqref{eq:Kherm} for the field $H_{\al\be}$ introduced in \eqref{eq:Kalbedef}. Taking into account \eqref{eq:Kalbe*}, the reality of $\ti{N}$ and the hermiticity of $H_{\al\be}$ translate into the reality of $N^{\al}$ and $\ti{q}_{\al\be}$.

All of the above equations can now be derived through the canonical (complex non-holomorphic) action
\beq \label{eq:Scancomp}
S = \int \ed^4 x \[ \frac{1}{\iu} \( E_i^{\al} \dot{A}_{\al}^i - \te^i \cG_i - N^{\al} \cD_{\al} \) - \ti{N} \cH - \xi^{(1)}_{\al\be} \, \cC_{(1)}^{\al\be} \]  \, , \hspace{1cm} \ti{N}, N^{\al}, \xi_{(1)}^{\al\be} \in \Rs \, ,
\eeq
where
\bea
\cH & \equiv & \cH(\bm{\psi}(E,B))  \, , \\
\cD_{\al} & := & \vep_{\al\be\ga} E_i^{\be} B^{i\ga} - A_{\al}^i \cG_i \equiv E_i^{\be} \( \pa_{\al} A_{\be}^i - \pa_{\be} A_{\al}^i \) - A_{\al}^i \pa_{\be} E_i^{\be} \, , \label{eq:Diffcomp} \\
\cG_i & := & D_{\al} E_i^{\al} \, , \label{eq:Gausscomp} \\
\cC_{(1)}^{\al\be} & := & {\rm Im} [ \ti{q}^{\al\be} ] \equiv {\rm Im} \[ \cH^{ij} E_i^{\al} E_j^{\be} \] \, , \label{eq:Calbedef}
\eea
by varying with respect to the complex fields $A_{\al}^i, \bar{A}_{\al}^i, E_i^{\al}, \bar{E}_i^{\al}, \te^i$ and the real fields $\ti{N}, N^{\al}, \xi_{(1)}^{\al\be}$. Again, the Lagrange multiplier $\xi^{(1)}_{\al\be}$ only appears in extra decoupled equations, the ones of the conjugate fields $\bar{A}_{\al}^i$ and $\bar{E}_i^{\al}$. In the canonical context there are no time-derivatives in the constraints, so trivial boundary conditions imply $\xi^{(1)}_{\al\be} = 0$. Note, also, that we have chosen to express the reality constraint on the 3-metric density through its inverse matrix in \eqref{eq:Calbedef}.

In the case of GR \eqref{eq:HGR}, we recover the (de-densitized) Ashtekar Hamiltonian constraint and the reality constraint on the electric fields 
\beq
\cH_{\rm GR} = \la + \de_{ij} E_{\al}^i B^{j\al} = 0 \, , \hspace{1cm}  \cC_{(1),\rm GR}^{\al\be} = {\rm Im}\[ \de^{ij} E_i^{\al} E_i^{\be} \] = 0 \, ,
\eeq
respectively. In particular, the latter constitutes six real constraints on the eighteen real fields in $E_i^{\al}$, so it admits a solution in terms of twelve undetermined real fields $e_{\al}^I \in \Rs$ 
\beq \label{eq:C1solGR}
E_i^{\al} = \vep^{\al\be\ga} \( \frac{1}{2}\, \vep_{ijk} e_{\be}^j e_{\ga}^k - \iu \de_{ij} e_{\be}^0 e_{\ga}^j \) \, ,
\eeq
since it leads indeed to the real 3-metric \eqref{eq:qNdef}
\beq
q_{\al\be} = \et_{IJ}\, e^I_{\al} e^J_{\be} \, .
\eeq
Thus, $N$, $N^{\al}$ and $e_{\al}^I$ together form the sixteen components of a vierbein 1-form
\beq \label{eq:eIofE}
e^I = \( - \frac{1}{6\sqrt{q}}\,N \vep^I_{\,\,\,JKL} \, \vep^{\al\be\ga}\, e_{\al}^J e_{\be}^K e_{\ga}^L + N^{\al} e_{\al}^I \) \ed t + e^I_{\al} \ed x^{\al} \, ,
\eeq
which is then related to the Urbantke metric \eqref{eq:UrbADM} through $g = \et_{IJ} \, e^I \otimes e^J$. In particular, inserting the expression \eqref{eq:C1solGR} back inside the action \eqref{eq:Scancomp}, one recovers the self-dual Palatini-Host action \eqref{eq:SPH} in terms of the vierbein \eqref{eq:eIofE}.

\subsection{Dirac algorithm and reality constraints}

In the standard cases where the action is real, or purely holomorphic, one can employ the Dirac algorithm for generating the whole constraint surface of the theory. In the present case, where the action \eqref{eq:Scancomp} is complex non-holomorphic, this option is unfortunately not applicable, or at least not for dealing with the reality constraint sector of the full constraint surface. 

To understand this point, we start by reminding that time evolution is given by the equations of the holomorphic theory \eqref{eq:dtA} and \eqref{eq:dtE}, because of the equations of motion of the conjugate fields. As a result, the evolution of some field/observable $X$ can be expressed by the standard formula
\beq \label{eq:evolHPB}
\dot{X} = - \{ \Hs_h, X \}_h \, , \hspace{1cm} \Hs_h := \int \ed^3 x \[ \te^i \cG_i + N^{\al} \cD_{\al} + \iu \ti{N} \cH \] \, ,
\eeq
where $\{ \cdot, \cdot \}_h$ is the holomorphic Poisson bracket \eqref{eq:holPB} and $\Hs_h$ is the holomorphic Hamiltonian. Note that considering the full Poisson bracket, i.e. including also the conjugate derivatives, is not an option, because this would lead to wrong evolution equations, since there is no $\bar{E}_i^{\al} \dot{\bar{A}}_{\al}^i$ term in the canonical action. Equation \eqref{eq:evolHPB} already shows that, in contrast with the usual case, the commutation relations among reality constraints are irrelevant for their conservation, since they simply do not appear there. As a result, the corresponding Lagrange multipliers cannot be determined through the Dirac algorithm process. This is consistent with a fact we already know, that the Lagrange multipliers of the reality constraints are instead determined by the equations of motion of the conjugate fields. These fields do not appear in \eqref{eq:evolHPB}, so their crucial equations are not taken into account in the Dirac algorithm.

Let us be more explicit by considering an example, the conservation of $\cC_{(1)}^{\al\be} = 0$. We want to show how the naive application of the Dirac algorithm can lead to mistakes, so we will ignore the equations of motion of the conjugate fields that eliminate the reality constraint contributions in the time evolution equations. We therefore have, instead of \eqref{eq:evolHPB}, 
\beq \label{eq:evolHPB2}
\dot{X} = - \{ \Hs, X \}_h \, , \hspace{1cm} \Hs := \int \ed^3 x \[ \te^i \cG_i + N^{\al} \cD_{\al} + \iu \( \ti{N} \cH + \xi^{(1)}_{\al\be} \cC_{(1)}^{\al\be} \) \] \, .
\eeq
Computing the corresponding evolution of $\cC_{(1)}^{\al\be}$, and taking into account the covariance of the constraint under 3-diffeomorphisms and SO(3,$\Cs$) transformations, we now find
\beq
\dot{\cC}_{(1)}^{\al\be}(x) \approx - \iu\left \{ \int \ed^3 y \[ \ti{N} \cH + \xi^{(1)}_{\ga\de} \, \cC_{(1)}^{\ga\de} \](y), \cC_{(1)}^{\al\be}(x) \right\} \, ,
\eeq
where ``$\approx$'' is the usual weak equality, i.e. up to terms that vanish on the constraint surface. Imposing that this expression be weakly zero leads to a local equation of the form
\beq \label{eq:DAex}
C_{(2)}^{\al\be} \ti{N} + D^{\al\be,\ga\de} \xi^{(1)}_{\ga\de} = 0 \, .
\eeq
There are now two possible scenarios: either the operator $D^{\al\be,\ga\de}$ is invertible, in which case this equation determines the Lagrange multiplier $\xi^{(1)}_{\al\be} = - (D^{-1})_{\al\be,\ga\de} C_{(2)}^{\ga\de} \ti{N}$, or it is not, in which case part of $\xi^{(1)}_{\al\be}$ is undetermined and we obtain extra secondary constraints, since the lapse function cannot vanish $\ti{N} \neq 0$. In the limit case where $D^{\al\be,\ga\de} = 0$, we obtain six secondary constraints $C_{(2)}^{\al\be} = 0$. If we now reintroduce the equations of motion of the conjugate fields, their effect is $\xi^{(1)}_{\al\be} = 0$ and thus replaces \eqref{eq:DAex} with $\ti{N} C_{(2)}^{\al\be} = 0$, meaning that we have a secondary constraint, {\it independently} of $D^{\al\be,\ga\de}$. In fact, here we will deal with theories where $D^{\al\be,\ga\de} \neq 0$, so the naive application of the Dirac algorithm would lead to erroneous conclusions.

Thus, reality constraints are a class of their own, on top of the first and second class ones in the holomorphic sector. In particular, they do not contribute to the Dirac bracket made out of second-class constraints. Lacking at present a more general formalism than the Dirac one that could encompass reality constraints, we can only ensure their conservation by computing explicitly their time derivative under the holomorphic flow and generating new constraints until they are all conserved.

Finally, let us note that not imposing $\xi^{(1)}_{\al\be} = 0$ can be interpreted as working with the corresponding ``real'' theories, i.e. where one includes also the conjugate kinetic sector in the action. Indeed, in that case the equations of motion of the conjugate fields are simply the conjugate of the holomorphic fields, instead of $\xi^{(1)}_{\al\be} = 0$, the evolution is given by \eqref{eq:evolHPB2} and the reality constraints enter the Dirac algorithm just like any other constraint. Thus, as already observed in section \ref{sec:actamb}, it is really the ``holomorphic + reality constraints'' form of the action that leads to this special status for the reality constraints, because of the asymmetry between the holomorphic and anti-holomorphic sectors. Incidentally, we will see that the only cases where $D^{\al\be,\ga\de} = 0$ are GR and CSDG, because then $\ti{q}^{\al\be}$ is $\de^{ij} E_i^{\al} E_j^{\be}$ and $\de^{ij} B_i^{\al} B_j^{\be}$, respectively, so $\cC_{(1)}^{\al\be} := {\rm Im}[\ti{q}^{\al\be}]$ only depends on one of the two canonical fields and thus commutes with itself trivially. As a result, for the {\it real} theories, only GR and CSDG have six extra constraints $C_{(2)}^{\al\be} = 0$, meaning that one can expect more degrees of freedom for the other theories.

\subsection{Towards the full constraint surface} \label{sec:towfullCS}

The situation described in the previous subsection forces us to generate the constraints in the brute force manner at the level of the equations of motion, i.e. by computing the time evolution of each one of them using \eqref{eq:dtA} and \eqref{eq:dtE} and then simplifying the result using the constraints. We start with the holomorphic constraints and find that they are already conserved on the constraint surface
\beq
\dot{\cH}, \dot{\cD}_{\al}, \dot{\cG}_i \approx 0 \, .
\eeq
This is to be expected, since these are the constraints generating the local symmetries of the holomorphic theory and time evolution is the one of the holomorphic theory. For the reality constraints, we first note that $\ti{N}$ and $N^{\al}$ are not determined by the equations of motion, because they are Lagrange multipliers in \eqref{eq:Scancomp}, so we can always simply choose them to be real. On the other hand, $\cC_{(1)}^{\al\be} = 0$ is a phase space constraint, so we have to compute its time derivatives. But this means that we have to compute the time-derivatives of the 3-metric, i.e. the ``geometrodynamics'' of the theory. This was already computed in \eqref{eq:qtidotgen*}.
Thus, to obtain the conservation of the constraint $\dot{\cC}_{(1)}^{\al\be} \approx 0$ one must impose a secondary one
\beq \label{eq:C2gen}
\cC_{(2)}^{\al\be} := {\rm Re} \[ E_i^{(\al} \vep^{\be)\ga\de} D_{\ga} E_{\de}^i - C^{ij,m}_{kl} E^{-1} E_i^{\al} E_j^{\be} E_m^{\ga} D_{\ga} \psi^{kl} \] = 0 \, .
\eeq 
Implementing this new constraint at the action level leads to the modification of \eqref{eq:Scancomp}
\beq \label{eq:Scancomp2}
S \to \int \ed^4 x \[ \frac{1}{\iu} \( E_i^{\al} \dot{A}_{\al}^i - \te^i \cG_i - N^{\al} \cP_{\al} \) - \ti{N} \cH - \xi^{(1)}_{\al\be} \, \cC_{(1)}^{\al\be} - \xi^{(2)}_{\al\be} \, \cC_{(2)}^{\al\be} \]  \, ,
\eeq
where $\xi^{(2)}_{\al\be} \in \Rs$. Again, the equations of the conjugate fields make $\xi^{(1)}_{\al\be}$ and $\xi^{(2)}_{\al\be}$ decouple from the other equations, thus preserving the holomorphic time evolution equation \eqref{eq:evolHPB}. Moreover, the equations for $\xi^{(1)}_{\al\be}$ and $\xi^{(2)}_{\al\be}$ are linear spatial differential equations whose unique solution, assuming trivial boundary conditions, is $\xi_{\al\be}^{(1,2)} = 0$. 

We must now determine whether $\cC_{(2)}^{\al\be}$ is also conserved, or whether it leads to a tertiary constraint. For the general case the task is complicated by the involved expression in \eqref{eq:C2gen}. In particular, note that further time derivatives will bring further derivatives of the $\cH(\bm{\psi})$ function. Therefore, for generic functions, one would expect the process to keep generating new independent combinations of the canonical fields and, consequently, an infinite series of constraints. This would mean that the reality constraints are not compatible with the holomorphic dynamics, i.e. they over-determine the system, as one would expect for generic complex non-holomorphic actions. There remains a possibility that for some choice of $\cH(\bm{\psi})$ the reality conditions are compatible with the dynamics. We now analyse the case of modified dynamics for the four-parameter family \eqref{eq:tcHext}. In that case, given \eqref{eq:dotqext} and the fact that $N$ is real, the secondary constraint is
\beq
{\rm Im} \, K_{\al\be} = 0 \, ,
\eeq
where the extrinsic curvature is given by \eqref{eq:SKdef}, which is therefore much simpler than the general case \eqref{eq:C2gen}.

\subsection{Degree of freedom count for the minimal polynomial modified theories} \label{sec:dofcount}

The analysis in Section \ref{sec:polynomial} provides us with everything we need in order to count the degrees of freedom for the polynomial modified theories \eqref{eq:tcHext}.

\subsubsection{The case $\De = 0$}

If $\De = 0$, then the kinetic term of the effective source \eqref{eq:Teff} vanishes. As already noted, the corresponding theories are GR ($\la_S, \la_B = 0$) and CSDG ($\la_E, \la_S = 0$). In both cases the effective potential $V$ in \eqref{eq:TrVofEB} becomes a constant, meaning that the source is a pure cosmological constant. If that constant is chosen to be real, then all terms after $\dot{K}_{\al\be}$ in \eqref{eq:dotK} are real, so the reality of $q_{\al\be}$ and $K_{\al\be}$ is conserved automatically and there is no need for tertiary constraints. The degree of freedom count then goes as follows. We have 26 real constraints $\cH, \cD_{\al}, \cG_i, \cC_{(1)}^{\al\be}, \cC_{(2)}^{\al\be} = 0$ but, because of \eqref{eq:modconstrGR} with $\ro = {\rm const}.$ real and $\cP_{\al} = 0$, only 22 of them are independent. To these 22 constraints we can then add 10 more real gauge-fixing conditions by choosing the 10 free real fields in $N, N^{\al}, \te^i$. This makes a total of 32 independent real equations for the 36 real canonical fields in $A_{\al}^i$ and $E_i^{\al}$, so we find 4 real canonical degrees of freedom. 

Now although CSDG also satisfies the vacuum Einstein equations for the real 4-metric $g_{\mu\nu}$, this does not mean that it is GR in disguise. Indeed, as already shown in section \ref{sec:CSDgravity} at the Lagrangian level, the reality constraints actually force the spin connection to be (A)dS. At the canonical level, this can be understood by noting that both the 3-metric and the extrinsic curvature depend exclusively on $A_{\al}^i$
\beq \label{eq:qtKCSDG}
\ti{q}^{\al\be} = \de_{ij} B^{i\al} B^{j\be} \, , \hspace{1cm} K_{\al\be} = \frac{\iu}{2q} \, q_{\al\ga\de} \de_{ij} B^{i\ga} \na_{\be} B^{j\de} \, ,
\eeq
so the reality constraints (12 real equations) completely determine $A_{\al}^i$ (18 real components $-$ 6 pure-gauge) and only $A_{\al}^i$. As a result, the only admissible solution for $g_{\mu\nu}$ is (A)dS. All the degrees of freedom are therefore in $E_i^{\al}$, which is completely independent of $q_{\al\be}$. A convenient way to arrange these degrees of freedom is in the complex tensor field $h^{\al\be} := q^{-1} B^{i\al} E_i^{\be}$. Since the reality constraints completely determine $A_{\al}^i$ and only $A_{\al}^i$, the holomorphic constraints constrain only $h^{\al\be}$ and make it a symmetric-traceless divergence-free tensor with respect to the geometry $q_{\al\be}$
\beq
h^{\al\be} = h^{\be\al} \, , \hspace{1cm} q_{\al\be} h^{\al\be} = 0 \, , \hspace{1cm} \na_{\al} h^{\al\be} = 0 \, .
\eeq 
Further noting that \eqref{eq:qtKCSDG} implies that $\na_{\al} B^{i\be}$ is completely determined by $K_{\al\be}$
\bea
\na_{\ga} B^{i\be} & \equiv & B_{i\al} \de_{jk} B^{j\al} \na_{\ga} B^{k\be} \equiv B_{i\al} \de_{jk} \( B^{j(\al} \na_{\ga} B^{k\be)} + B^{j[\al} \na_{\ga} B^{k\be]} \) \nn \\
 & = & B_{i\al} \( \frac{1}{2} \, \na_{\ga} \ti{q}^{\al\be} - \iu q q^{\al\be\de} K_{\de\ga} \) \equiv - \iu B_{i\al} q q^{\al\be\de} K_{\de\ga} \, ,
\eea 
the evolution equations \eqref{eq:dtcB} and \eqref{eq:dtcE} translate into the linear equation for $h^{\al\be}$
\beq \label{eq:dtEBB}
\( \pa_t - \Lie_N \) h^{\al\be} = N \[ K^{\be}_{\ga} h^{\al\ga} - 2 K h^{\al\be} - q^{\al\be} K_{\ga\de} h^{\ga\de} - \iu q^{\be\ga}_{\,\,\,\,\,\,\,\de} \na_{\ga} h^{\al\de} \] \, .
\eeq
So $E_i^{\al}$ carries four degrees of freedom, but arranged in a complex tensor field satisfying a chiral first-order evolution equation \eqref{eq:dtEBB}. In contrast, in GR the reality constraints depend on both canonical fields 
\beq
\ti{q}^{\al\be} = \de^{ij} E_i^{\al} E_j^{\be} \, , \hspace{1cm} K_{\al\be} = \frac{\iu}{2q} \, q_{\al\ga\de} \de^{ij} E_i^{\ga} \na_{\be} E_j^{\de} \, ,
\eeq
and therefore constraint one tensor combination in each one of them. This leaves us again with four degrees of freedom, but now arranged in a real tensor field $q_{\al\be}$ obeying non-chiral second-order evolution equations.

\subsubsection{The case $\De \neq 0$}

This is the case of most interesting modifications. If $\De \neq 0$, then the right-hand side of \eqref{eq:dotK} is no longer automatically real in general. The effective energy and momentum densities $\ro$ and $\cP_{\al}$ are real, because of \eqref{eq:modconstrGR}, but not the effective stress $\cS_{\al\be}$. From \eqref{eq:effS} we thus see that we require a tertiary constraint
\beq\label{eq:tertiary}
\cC^{(3)}_{\al\be} := {\rm Im} \[ \De \( S_{\al\ga\de} S_{\be}^{\,\,\,\ga\de} - q_{\al\be} S_{[\ga\de]\ep} S^{\ga\de\ep} \) \] = 0 \, ,
\eeq
which we can also express more neatly, yet redundantly, in 4d language \eqref{eq:Teff}
\beq
{\rm Im} \[ T^{\rm eff}_{\mu\nu} - \frac{1}{2}\, g_{\mu\nu} T^{\rm eff} \] \propto {\rm Im} \[ \frac{\De}{4}\, \cH_{ik} \cH_{jl} D_{\mu} \psi^{ij} D_{\nu} \psi^{kl} + g_{\mu\nu} V \] = 0 \, .
\eeq 
However, from the degree of freedom count of the $\De = 0$ case, which already has the minimum field content of a single graviton, we already know that tertiary constraints leave no room for degrees of freedom at all. Thus, the $\De \neq 0$ theories are overdetermined, meaning that they only admit the maximally symmetric solution $E_i^{\al} = c \de_{ij} B^{j\al}$, for some constant $c$. We therefore see that the discriminant parameter $\De$ determines whether there are degrees of freedom or not, after reality conditions are imposed. Interestingly, this parameter also controls the Poisson bracket of the 3-metric with itself
\beq \label{eq:qtiselfcomm} 
\{ q^{\al\be}, q^{\ga\de} \} = - \frac{\De}{2} \[ \vep^{\al\ga\ep} S_{\ep}^{\,\,\,\be\de} + \vep^{\al\de\ep} S_{\ep}^{\,\,\,\be\ga} + \vep^{\be\ga\ep} S_{\ep}^{\,\,\,\de\al} + \vep^{\be\de\ep} S_{\ep}^{\,\,\,\al\ga} \] \, , 
\eeq
Given the symmetries of this tensor density and the fact that 
\beq
q_{\ga\de}\, \{ q^{\al\ga}, q^{\be\de} \} = 0 \, ,
\eeq
we can invert the relation  
\beq\label{eq:S-comm}
S_{\ga}^{\,\,\,\al\be} = -\frac{1}{2\De} \, \vep_{\ga\de\ep} \, \{ q^{\al\de}, q^{\be\ep} \} \, ,
\eeq
thus providing a ``symplectic'' interpretation for $S_{\ga}^{\,\,\,\al\be}$, in contrast to the geometric interpretation of $K_{\al\be}$. Therefore, for $\De \neq 0$ the 3-metric cannot even be turned into a canonical variable through some canonical transformation, as this would require $\{ q^{\al\be}, q^{\ga\de} \} = 0$. In the quantum theory, where the Poisson bracket is replaced by the commutator, this would imply that one cannot measure all the components of the 3-metric without some minimal uncertainty. 

We thus see that in the case $\De\not=0$ the 3-metric components do not Poisson commute with themselves, and this is linked to the appearance of the non-real extra terms in the evolution equations for the 3-metric and its time derivative. It is thus possible that this is just a signal that the 3-metric is not the right variable to require to be real. However, as we already discussed, there appears no other known way to impose the Lorentzian reality conditions.

\section{Conclusion}

In this paper we have performed a detailed canonical analysis of the minimal polynomial modifications of GR, which are theories given by the polynomial canonical action
\beq \label{eq:ScanMP}
\hat{S} = \int \ed^4 x \[ \frac{1}{\iu} \( E_i^{\al} \dot{A}_{\al}^i - \te^i \cG_i - N^{\al} \cD_{\al} \) - \hat{N} \hat{\cH} - \xi^{(1)}_{\al\be} \, \cC_{(1)}^{\al\be} \]  \, , \hspace{1cm} \hat{N}, N^{\al}, \xi^{(1)}_{\al\be} \in \Rs \, ,
\eeq
with \eqref{eq:MPhatH}, \eqref{eq:Diffcomp}, \eqref{eq:Gausscomp} and $\cC_{(1)}^{\al\be} := {\rm Im} \, \ti{q}^{\al\be}$ with the 3-metric given by \eqref{eq:MPtiq}. 

Our first result is the complete analysis of the geometrodynamics of these theories, prior to any reality conditions being imposed. We have seen that the modification is controlled by the discriminant $\De$. The case $\De=0$ is that of GR and constrained Self-Dual Gravity. It is only in this case that the field equations of the theory can be written in a closed metric form. In the case $\De\not=0$, we have obtained an effective geometric description of the modifications, as Einstein equations with the effective stress-energy tensor \eqref{eq:Teff}. Importantly, the results obtained indicate that the essentially modified $\De\not=0$ theories are non-metric: They cannot be described as dynamical theories of solely a 4-dimensional metric, at least not with second order field equations. The question as to the geometrical interpretation of the modified theories, at least in the cases of Riemannian and split signatures when these modifications do make sense, remains open. It is possible that the right way to approach this question is via the link between these modifications and the higher-dimensionsional diffeomorphism-invariant theories of the type considered in  \cite{Krasnov:2016wvc}, \cite{Herfray:2016azk}. Let us also remark that the question of spin-foam type quantisation, see e.g. \cite{Perez:2012wv} for a review of the spin-form quantum gravity, of the Euclidean signature modified gravity theories we studied in this paper remains an interesting open question.

Our second main result is the analysis of the compatibility of the reality conditions requiring that the Urbantke metric \eqref{eq:Urbantke} is real and the evolution equations of the modified theories. We have seen that in the modified case these are not in general compatible in the sense that the requirement of the time conservation of the secondary reality conditions (stating that the time derivative of the 3-metric is real) produces tertiary constraints \eqref{eq:tertiary}. We have also seen that the non-vanishing of the tertiary constraints is directly linked to the tensor $S_\alpha{}^{\beta\gamma}$, which is in turn linked, see \eqref{eq:S-comm}, to the Poisson bracket of the 3-metric with itself. Our results thus appear to indicate that the 3-metric is not anymore a good canonical variable to be used in the modified case. Indeed, when modifications are present its components no longer Poisson commute, and thus no longer provide a useful set of coordinates on the corresponding phase space. As the result, imposing that these functions on the phase space are real does not produce constraints that are compatible with the modified dynamics. Thus, our results still leave open the possibility that there exist some other functions on the phase space of the complexified modified theories that can meaningfully be required to be real. It is however not at all clear how to go about searching for such functions. If anything, geometry should provide guidance. This makes the question of a geometric interpretation of the modified theories the most important open problem in this subject.

\acknowledgments

EM is supported by a Consolidator Grant of the European Research Council (ERC-2015-CoG grant 680886).

\appendix

\section{From Plebanski to Einstein-Hilbert} \label{app:toEH}

In the case of GR \eqref{eq:HGR}, including the reality constraints \eqref{eq:Sc}, the equations of motion of $\psi^{ij}$, $\ch$, $\ch^{ij}$
\beq \label{eq:LMEOM}
\frac{\iu}{2}\, B_i \we B_j = \ph \de_{ij}  \, ,   \hspace{1cm} \de^{ij} B_i \we B_j + \de^{ij} \bar{B}_i \we \bar{B}_j = 0 \, , \hspace{1cm} B_i \we \bar{B}_j = 0 \, ,
\eeq
form 22 real equations that reduce the 38 real variables in $B_i$ and $\ph$ to the 16 components of a real vierbein $e^I$
\beq
B_i = \frac{1}{8\pi G} \[ - \iu\, \de_{ij} e^0 \we e^j + \frac{1}{2}\, \vep_{ijk} e^j \we e^k \] \, , \hspace{1cm} \ph = \frac{1}{\( 8 \pi G \)^2} \, e^0 \we e^1 \we e^2 \we e^3 \, .
\eeq
Inserting this back inside the action then leads to the self-dual Palatini-Holst action
\beq \label{eq:SPH}
S = \frac{1}{16 \pi G} \int \[ \( \frac{1}{2}\, \vep_{IJKL} \, e^K \we e^L + \iu\, e_I \we e_J \) \we F^{IJ} - \frac{\La}{12}\, \vep_{IJKL} \, e^I \we e^J \we e^K \we e^L \] \, ,
\eeq
where
\beq \label{eq:FIJ}
F^{IJ} := \ed A^{IJ} + A^I_{\,\,\,K} \we A^{KJ}  \, , \hspace{1cm} A^{0i} := {\rm Im} [A^i] \, , \hspace{1cm} A^{ij} := - \vep^{ijk} \,{\rm Re}[A^k] \, , 
\eeq
are the curvature 2-forms of the real spin connection $A^{IJ}$ 1-forms and $\La := \la / (8\pi G)$ is the cosmological constant. One can then integrate out the spin connection, i.e. consider its equation of motion
\beq
e^{[I} \we \ced e^{J]} - \frac{\iu}{2}\, \vep^{IJ}_{\,\,\,\,\,\,KL} \, e^K \we \ced e^L = 0 \, , \hspace{1cm} \ced e^I := \ed e^I + A^I_{\,\,\,J} \we e^J \, ,
\eeq
which is solved by the torsion-free spin connection
\beq
A^{IJ} :=  e^{I\nu} \pa_{[\mu} e_{\nu]}^J - e^{J\nu} \pa_{[\mu} e_{\nu]}^I - e_{K \mu} e^{I\nu} e^{J\ro} \pa_{[\nu} e_{\ro]}^K \, ,
\eeq
and insert this back inside \eqref{eq:SPH} to obtain, after some work, the Einstein-Hilbert action for the real metric $g := \et_{IJ}\, e^I \otimes e^J$.

\section{Reality constraints and the Urbantke metric} \label{app:RCtometric}

We start by noting that the volume density $\sqrt{-g}$ of the Urbantke metric \eqref{eq:Urbantke} is precisely the combination that is made real by the scalar reality constraint ${\rm Re} \[ \cH^{ij} B_i \we B_j \] = 0$. We therefore only need to show that the tensor reality constraint $B_i \we \bar{B}_j = 0$ is equivalent to $g_{\mu\nu}$ being real, up to a possibly complex overall factor. We start by foliating space-time $x^{\mu} \to \{ t, x^{\al} \}$ and defining
\beq
E_i^{\al} := \frac{1}{2}\, \vep^{\al\be\ga} B_{i\be\ga} \, ,  \hspace{1cm} H_{\al\be} := \iu B_{it\al} E_{i\be} \, , 
\eeq
where $E_{\al}^i$ is the inverse matrix of $E_i^{\al}$. With this, the tensor reality constraint $B_i \we \bar{B}_j = 0$ becomes the hermiticity condition
\beq \label{eq:RCH}
H_{\al\be} = \bar{H}_{\be\al} \, .
\eeq
We can then write
\beq
H_{\al\be} \equiv Q^{-1/2} Q_{\al\be} + \iu \vep_{\al\be\ga} N^{\ga} \, , \hspace{1cm} Q_{\al\be} \equiv Q_{\be\al} \, , \hspace{1cm} Q := \det Q_{\al\be} \, ,
\eeq
so that $Q_{\al\be}$ and $N^{\al}$ have zero density weight and \eqref{eq:RCH} becomes simply
\beq \label{eq:RCH2}
{\rm Im} \, Q_{\al\be} = {\rm Im} \, N^{\al} = 0 \, .
\eeq
With these, the line-element of the Urbantke metric \eqref{eq:Urbantke} takes the form
\beq
g \propto - \ed t^2 + Q_{\al\be} \( \ed x^{\al} + N^{\al} \ed t \) \( \ed x^{\be} + N^{\be} \ed t \) \, ,
\eeq
which is therefore real indeed. Conversely, the reality of this metric, up to an overall factor, implies \eqref{eq:RCH2} and thus the tensor reality constraint $B_i \we \bar{B}_j = 0$.

\bibliographystyle{JHEP}
\bibliography{mybib}

\end{document}

%% file: mydefs.tex
\newcommand{\beq}{\begin{equation}}
\newcommand{\eeq}{\end{equation}}
\newcommand{\bea}{\begin{eqnarray}}
\newcommand{\eea}{\end{eqnarray}}
\newcommand{\ben}{\begin{enumerate}}
\newcommand{\een}{\end{enumerate}}

\newcommand{\pa}{\partial}

\newcommand{\na}{\nabla}

\newcommand{\ed}{{\rm d}}
\newcommand{\ced}{{\rm D}}
\newcommand{\we}{\wedge}

\newcommand{\Lie}{{\cal L}}
\newcommand{\Ord}{{\cal O}}

\newcommand{\Tr}{{\rm Tr}}
\newcommand{\ti}{\tilde}

\newcommand{\Rs}{\mathbb{R}}
\newcommand{\Cs}{\mathbb{C}}

\newcommand{\Hs}{\mathbb{H}}

\renewcommand\({\left(}
\renewcommand\){\right)}
\renewcommand\[{\left[}
\renewcommand\]{\right]}

\newcommand{\nn}{\nonumber}

\newcommand{\al}{\alpha}
\newcommand{\be}{\beta}
\newcommand{\ga}{\gamma}
\newcommand{\Ga}{\Gamma}
\newcommand{\de}{\delta}
\newcommand{\De}{\Delta}
\newcommand{\ep}{\epsilon}
\newcommand{\vep}{\varepsilon}
\newcommand{\ze}{\zeta}
\newcommand{\et}{\eta}
\newcommand{\te}{\theta}

\newcommand{\ka}{\kappa}
\newcommand{\la}{\lambda}
\newcommand{\La}{\Lambda}
\newcommand{\ro}{\rho}
\newcommand{\si}{\sigma}

\newcommand{\ta}{\tau}
\newcommand{\ph}{\phi}

\newcommand{\ch}{\chi}

\newcommand{\cH}{{\cal H}}
\newcommand{\cE}{{\cal E}}

\newcommand{\cP}{{\cal P}}
\newcommand{\cS}{{\cal S}}

\newcommand{\cD}{{\cal D}}

\newcommand{\cB}{{\cal B}}
\newcommand{\cC}{{\cal C}}
\newcommand{\cG}{{\cal G}}

\newcommand{\iu}{{\rm i}}